\documentclass[aps, preprint, nofootinbib,preprintnumbers,eqsecnum,superscriptaddress,sort,calc]
{revtex4}
\pdfoutput=1

\usepackage{color}

\usepackage[
      colorlinks=true,
      linkcolor=blue,
      urlcolor=blue,
      filecolor=black,
      citecolor=red,
      pdfstartview=FitV,
      pdftitle={},
        pdfauthor={Oscar Dias, Ramon Masachs},
        pdfsubject={},
        pdfkeywords={},
        pdfpagemode=None,
        bookmarksopen=true
      ]{hyperref}

\usepackage[normalem]{ulem}
\usepackage{amsmath}
\usepackage{enumerate}
\usepackage{amsfonts}
\usepackage{yfonts}

\usepackage{subfigure}
\usepackage{psfrag}

\usepackage{epsfig}
\usepackage[utf8]{inputenc}
\usepackage{float}
\usepackage{graphicx}
\usepackage{cancel}
\usepackage{mathrsfs}
\usepackage{amssymb}
\usepackage{amsfonts}
\usepackage{amsmath}
\usepackage{slashed}
\usepackage{comment}

\usepackage{graphicx}
\usepackage{bm}

\def\eg{{\it e.g. }}
\def\ie{{\it i.e. }}

\def\({\left(}
\def\){\right)}
\def\[{\left[}
\def\]{\right]}
\def\<{\langle}
\def\>{\rangle}

\newcommand{\be}{\begin{equation}}
\newcommand{\ee}{\end{equation}}
\newcommand{\bea}{\begin{eqnarray}}
\newcommand{\eea}{\end{eqnarray}}
\newcommand{\bwt}{\begin{widetext}}
\newcommand{\ewt}{\end{widetext}}

\newcommand{\bi}{\begin{itemize}}
\newcommand{\ei}{\end{itemize}}
\newcommand{\ben}{\begin{enumerate}}
\newcommand{\een}{\end{enumerate}}
\newcommand{\bca}{\begin{cases}}
\newcommand{\eca}{\end{cases}}
\newcommand{\bln}{\begin{align}}
\newcommand{\eln}{\end{align}}
\newcommand{\bst}{\begin{split}}
\newcommand{\est}{\end{split}}

\begin{document}

\title {Charged black hole bombs in a Minkowski cavity}


\author{Oscar J. C. Dias}
\email{ojcd1r13@soton.ac.uk}
\affiliation{STAG research centre and Mathematical Sciences, University of Southampton, UK \vspace{1 cm}}

\author{Ramon Masachs}
\email{rmg1e15@soton.ac.uk}
\affiliation{STAG research centre and Mathematical Sciences, University of Southampton, UK \vspace{1 cm}}

\begin{abstract}
Press and Teukolsky famously introduced the concept of a black hole bomb system: a scalar field scattering a Kerr black hole confined inside a mirror undergoes superradiant amplification that keeps repeating due to the reflecting boundary conditions at the mirror. A similar charged black hole bomb system exists if we have a charged scalar field propagating in a Reissner-Nordstr\"om black hole confined inside a box. We point out that scalar fields propagating in such a background are unstable not only to superradiance but also to a mechanism known as the near-horizon scalar condensation instability. The two instabilities are typically entangled but we identify regimes in the phase space where one of them is suppressed but the other is present, and vice-versa (we do this explicitly for the charged but non-rotating black hole bomb). These `corners' in the phase space, together with a numerical study of the instabilities allow us to identify accurately the onset of the instabilities. Our results should thus be useful to make  educated choices of initial data for the Cauchy problem that follows the time evolution and endpoint of the instabilities. Finally, we use a simple thermodynamic model (that makes no use of the equations of motion) to find the leading order thermodynamic properties of hairy black holes and solitons that should exist as a consequence (and that should be the endpoint) of these instabilities. In a companion publication, we explicitly solve the Einstein-Maxwell-scalar equations of motion to find the properties of these hairy solutions at higher order in perturbation theory. 
\end{abstract}

\today

\maketitle

\tableofcontents



\section{Introduction}\label{section:Introduction}

The black hole bomb system was envisaged by Press and Teukolsky more than 4 decades ago \cite{Press:1972zz,Bardeen:1972fi}. This is a setup whereby  superradiant scalar waves continuously extract energy and angular momentum from a Kerr black hole that is placed  in the interior of a reflecting wall (mirror)\footnote{We will use the words `mirror', `box', `cavity', `wall', `surface layer' or even `cage' to refer to this mirror surface layer with relecting boundary conditions.}. This occurs as long as  the frequency $\omega$, azimuthal angular momentum $m_{\varphi}$ and Kerr angular velocity satisfy the superradiant condition $\omega<m_{\varphi}\Omega_H$ \cite{Zeldovich:1971,Starobinsky:1973,Teukolsky:1974yv} and until the confining box can no longer hold the radiation pressure \cite{Press:1972zz} (see \cite{Brito:2015oca} for a review).

The typical instability timescales of this black hole bomb system are known 
\cite{Cardoso:2004nk,Dolan:2012yt,Ferreira:2017tnc,Hod:2016rqd}. However, the time evolution and endpoint of the black hole bomb instability are still missing. Yet we have some hints of the answer from other systems that mimic this setup. Actually, these partner black hole bomb systems turn out to provide more natural reflecting boundary conditions. These are rotating black holes placed inside global AdS (initially discussed in \cite{Hawking:1999dp,Cardoso:2004hs}) and rotating black holes scattered by massive bosonic fields (\eg scalar, Proca), as initially proposed in \cite{Zouros:1979iw,Detweiler1980}. In the former case, the global AdS boundary is timelike (it is conformal to the Einstein static universe) and if we impose boundary conditions that do not deform this asymptotic boundary, there is no flux of energy and angular momentum: superradiant waves get confined and increasingly amplified after each reflection at the boundary. In the latter case, the mass of the bosonic field effectively provides a potential barrier that traps the bosonic waves near the horizon. From the superradiant studies in rotating AdS black holes it is conjectured that  superradiant instabilities should evolve following one of two possible scenarios \cite{Cardoso:2004hs,Dias:2011at,Dias:2013sdc,Cardoso:2013pza,Dias:2015rxy,Niehoff:2015oga,Choptuik:2017cyd}. Namely, it should evolve: 1) towards a singular solution reached in finite time which implies  cosmic  censorship  violation or, most probably, 2) towards the excitation of higher and higher $m_{\varphi}$ modes until the cosmic censorship conjecture is eventually violated. In the sense that as $m_{\varphi}\to \infty$, the small lengthscales that are reached should signal a breakdown of classical gravity \cite{Dias:2011at,Dias:2013sdc,Cardoso:2013pza,Dias:2015rxy,Niehoff:2015oga,Choptuik:2017cyd}. This suggests that the original  Press-Teukolsky black hole bomb system might have the same fate. However, a final answer is still awaiting for a full time evolution of the system. 

We can also have non-rotating superradiant black hole bomb systems, as long as the system has electric charge \cite{Herdeiro:2013pia,Hod:2013fvl,Degollado:2013bha,Hod:2014tqa,Li:2014gfg,Dolan:2015dha,Basu:2016srp,Ponglertsakul:2016wae,Ponglertsakul:2016anb,Sanchis-Gual:2015lje,Sanchis-Gual:2016tcm,Sanchis-Gual:2016ros,Hod:2016kpm,Fierro:2017fky,Li:2014xxa,Li:2014fna,Li:2015mqa,Li:2015bfa}. Indeed, a scalar field with charge $q$ scattering a charged black hole with chemical potential $\mu$ can also be superradiantly amplified if the wave frequency $\omega$ satisfies the bound $\omega< q\mu$.  For example, this can be the charged partner of Press-Teukolsky's system $-$ a Reissner-Nordstr\"om black hole placed inside a box with a complex scalar field \cite{Herdeiro:2013pia,Hod:2013fvl,Degollado:2013bha,Hod:2014tqa,Li:2014gfg,Dolan:2015dha,Sanchis-Gual:2015lje,Basu:2016srp,Ponglertsakul:2016wae,Ponglertsakul:2016anb,Sanchis-Gual:2016tcm,Sanchis-Gual:2016ros,Hod:2016kpm,Fierro:2017fky} (see also \cite{Li:2014xxa,Li:2014fna,Li:2015mqa,Li:2015bfa})$-$ or a Reissner-Nordstr\"om black hole in global AdS \cite{Bhattacharyya:2010yg,Basu:2010uz,Dias:2011tj,Gentle:2011kv,Dias:2016pma}. These charged systems are interesting on their own and because they can provide insights for the endpoint of the rotating superradiant instability.\footnote{In spite of the close connection between the rotating and static-charged superradiant systems there is also a fundamental distinction. In the latter case, $q$ is fixed once we choose the theory and thus the associated hairy black holes are not further unstable to superradiance (except if we introduce a second scalar field with charge $q_2$). On the other hand, the hairy black holes that emerge from the onset of the rotating superradiant instability have a particular $m_\varphi$ and are still unstable to superradiant modes with higher $m_\varphi$.} In such cases, the endpoint of the superradiant instability is either known or we have already good hints of what it might be. Indeed, in global AdS, the time simulations of \cite{Bosch:2016vcp,Arias:2016aig} confirmed that the endpoint of charged superradiance is one of the hairy black holes of \cite{Basu:2010uz,Dias:2011tj,Gentle:2011kv,Dias:2016pma}. Similarly, for charged superradiance on Reissner-Nordstr\"om black holes with a mirror, recent time simulations indicate that the endpoint is again a charged hairy black hole of Einstein-Maxwell theory \cite{Sanchis-Gual:2015lje,Sanchis-Gual:2016tcm,Sanchis-Gual:2016ros}. 

In this paper, and in a companion manuscript \cite{DiasMasachs}, we study the Reissner-Nordstr\"om black hole bomb system in more detail. We have two main goals: 1)  perform a detailed study of the linear mode instability problem (in this manuscript), and 2) find the regular and asymptotically flat charged hairy black holes of the theory that describe the endpoint of the instability (in the companion manuscript \cite{DiasMasachs}).

Previous literature already addressed several aspects of our linear mode stability problem (and even its nonlinear evolution) \cite{Herdeiro:2013pia,Hod:2013fvl,Degollado:2013bha,Hod:2014tqa,Li:2014gfg,Dolan:2015dha,Sanchis-Gual:2015lje,Basu:2016srp,Ponglertsakul:2016wae,Ponglertsakul:2016anb,Sanchis-Gual:2016tcm,Sanchis-Gual:2016ros,Hod:2016kpm,Fierro:2017fky}. Our aim is to complement and complete these studies in three main directions.
First we want to point out that the system is unstable to superradiance but also to the near-horizon scalar condensation instability \cite{Gubser:2008px,Hartnoll:2008vx,Dias:2010ma} (Section \ref{sec:RNinstabilityHeuristics}). The latter source of instability seems to have been missed in previous studies of the black hole bomb mirror system. Essentially, this instability is present because a scalar field can violate the Breitenl\"ohner-Freedman \cite{Breitenlohner:1982jf} bound associated with the AdS$_2\times S^2$ near-horizon geometry of the extremal Reissner-Nordstr\"om configuration\footnote{This is the very same instability that ignited the holographic AdS/condensed matter correspondence programme \cite{Gubser:2008px,Hartnoll:2008vx,Dias:2010ma}.}. Thus it has a different nature than the superradiant instability. These two instabilities are typically entangled. However, there are two limits (which we will identify) where they disentangle and expose their different nature: in units of the box radius, {\it large} caged Reissner-Nordstr\"om black holes only have the near-horizon instability, while {\it small} caged Reissner-Nordstr\"om black holes are unstable only to superradiance. 

Built on these considerations, we will then proceed to our second main aim (Section \ref{sec:QNM}). Namely, we find the onset surface for instability in the Reissner-Nordstr\"om$-$mirror system and key properties of the instability. To begin, we find the critical  scalar field charge, as a function of the horizon radius and chemical potential, above which the system is unstable. The exact identification of this onset surface is absent in previous literature and proves useful, for example, to identify good initial data for time evolution simulations. Our results are exact and considerably sharpen (and testify) the analytical bounds necessary for the existence of instability given in \cite{Herdeiro:2013pia,Hod:2016kpm}. For this exercise we use a scaling symmetry of the system to fix the mirror at radius $R=1$ (Section \ref{sec:Model}). Reissner-Nordstr\"om black holes are parametrized by two parameters, \eg the (dimensionless) horizon radius $R_+$ and the chemical potential $\mu$. The horizon must then be inside the mirror, \ie with radius $R_+<1$. The extremality condition $\mu=\mu_{ext}$  determines a 1-parameter family of Reissner-Nordstr\"om black holes that have zero temperature. The onset curve  $e_{onset}(R_+)$ $-$ to be displayed in Fig. \ref{fig:eonset_Rp_chp} $-$ describes the minimal scalar field (dimensionless) electric charge above which  extremal Reissner-Nordstr\"om black holes are unstable (non-extremal black holes are unstable above even higher charges $e_{onset}(R_+,\mu)$ that are also identified in Section \ref{sec:QNM}). We also study exhaustively the frequency spectrum of the instability. In particular, we  show that the most unstable modes are those described by a spherical harmonic with angular quantum number $\ell=0$, in agreement with the time domain analysis of \cite{Degollado:2013bha} (note that the frequency-domain study of \cite{Herdeiro:2013pia} just considered the $\ell=1$ harmonic).

Our third goal is to use the properties of the linear instability problem to infer the existence and properties of hairy black holes that should bifurcate (in a phase diagram of solutions of the theory) from the Reissner-Nordstr\"om family at the onset of the instability (Section \ref{sec:noninteracting}). Without the box, no-hair theorems forbid the existence of regular asymptotically flat solitons and black holes \cite{Ruffini:1971bza,Chrusciel:1994sn,Bekenstein:1996pn,Heusler:1998ua}. Essentially, this is because a linearized scalar field that extends all the way up to asymptotic infinity, when back-reacted on the gravitoelectric fields, sources terms that diverge logarithmically as the radius grows large (see Appendix \ref{App:HoHair}). However, this no-hair theorem can be evaded if we confine the scalar field inside a box. In these conditions it should be possible to have asymptotically flat solutions that are regular everywhere. Then the theory should have a hairy charged soliton (aka boson star)  that is the back-reaction of a normal mode in a box. Assuming this, we can use  a simple {\it non}-interacting thermodynamic model \cite{Basu:2010uz,Dias:2011at,Dias:2011tj,Cardoso:2013pza,Dias:2015rxy,Dias:2016pma} to predict the {\it leading} order thermodynamics (in a small mass and charge expansion) of the hairy black holes. This thermodynamic model assumes that the hairy black hole can be constructed placing a small Reissner-Nordstr\"om black hole on top of the soliton (aka boson star) of the theory, subject to the condition that they have the same chemical potential. The soliton is just a regular asymptotically flat horizonless solution with scalar hair that results from the back-reaction of a scalar field normal mode of a Minkowski box. Interestingly, this thermodynamic model does {\it not} make use of the equations of motion. Yet, it does yield the correct leading order thermodynamics of the charged hairy black hole. This will be explicitly confirmed in our companion manuscript \cite{DiasMasachs} where we find the charged hairy black holes solving explicitly  the equations of motion. To be clear, these are  regular asymptotically flat black hole solutions of the Einstein-Maxwell theory with a scalar field confined in a reflecting box.
 
\section{Reissner-Nordstr\"om black hole bomb system}\label{sec:Model}

Einstein-Maxwell theory is described by the action
\begin{align}\label{action}
S=\frac{1}{16 \pi G_4}\int\mathrm d^4 x\sqrt{g}\left({\cal R}-\frac 1 2F_{\mu\nu}F^{\mu\nu}\right),
\end{align}
where ${\cal R}$ is the Ricci scalar, $A$ is Maxwell's potential and $F=\mathrm d A$ is Maxwell's field strength. We fix Newton's constant as $G_4\equiv 1$. 
The Reissner-Nordstr\"om black hole (RN BH) family of solutions of this theory is a two-parameter family that we can take to be the horizon radius $r_+$ and the chemical potential $\mu=A(\infty)-A(r_+)$. In the gauge where $A(r_+)=0$ this solution is described  by 
\begin{align}\label{fieldansatz}
\mathrm d s^2=-f(r)\mathrm d t^2+g(r)\mathrm d r^2+r^2\mathrm d \Omega_2^2, \qquad A_{\mu}\mathrm d x^{\mu}=A(r)\mathrm d t, 
\end{align} 
with $\mathrm d \Omega_2^2$ being the metric for the unit 2-sphere and
\begin{equation}\label{eq:RNbackground}
f(r)=1-\frac{r_+}{r}\frac{2+\mu^2}{2}+\frac{r_+^2}{r^2}\frac{\mu^2}{2},\qquad g(r)=\frac{1}{f(r)},\qquad  A(r)=\mu\left(1-\frac{r_+}{r}\right).
\end{equation} 

 The ADM mass, electric charge, entropy, and temperature of the RN BH are 
\begin{equation}\label{RNthermo}
M=\frac{1}{4}(2+\mu^2)r_+\,,\qquad Q=\frac{1}{2}\,\mu\,r_+\,, \qquad S_H=\pi r_+^2\,, \qquad T_H L=\frac{1}{8\pi  r_+}(2-\mu^2)\,.
\end{equation}
Extremal RN BHs ($T_H=0$) have $\mu=\sqrt 2$, so RN BHs exist for $\mu\leq \sqrt 2$.

RN BHs are known to be linear mode stable to gravitoelectromagnetic perturbations but also to scalar field perturbations. However, we will be interested in placing an asymptotically flat RN BH inside a box $\Sigma$ of radius $r=L>r_+$. In this case, 
 a test scalar field $\phi$ with charge $q$ and mass $m$ can drive the system unstable. Such a complex scalar field obeys the Klein-Gordon equation,
\begin{equation}\label{KG}
D_\mu D^\mu\phi-V^\prime(|\phi |^2)\phi=0\,,
\end{equation}
where $D_{\mu}=\nabla_{\mu}-i q A_{\mu}$ is the gauge covariant derivative. 
We consider potentials that have the expansion $V(\eta)=m^2 \eta+O(\eta^2)$ such that $m$ is the scalar field mass. Most of our physical discussions will apply to a generic case $m$. However, for concreteness, our numerical results will restrict to the massless case $m=0$. 

\begin{figure}[t]
\centerline{
\includegraphics[width=.48\textwidth]{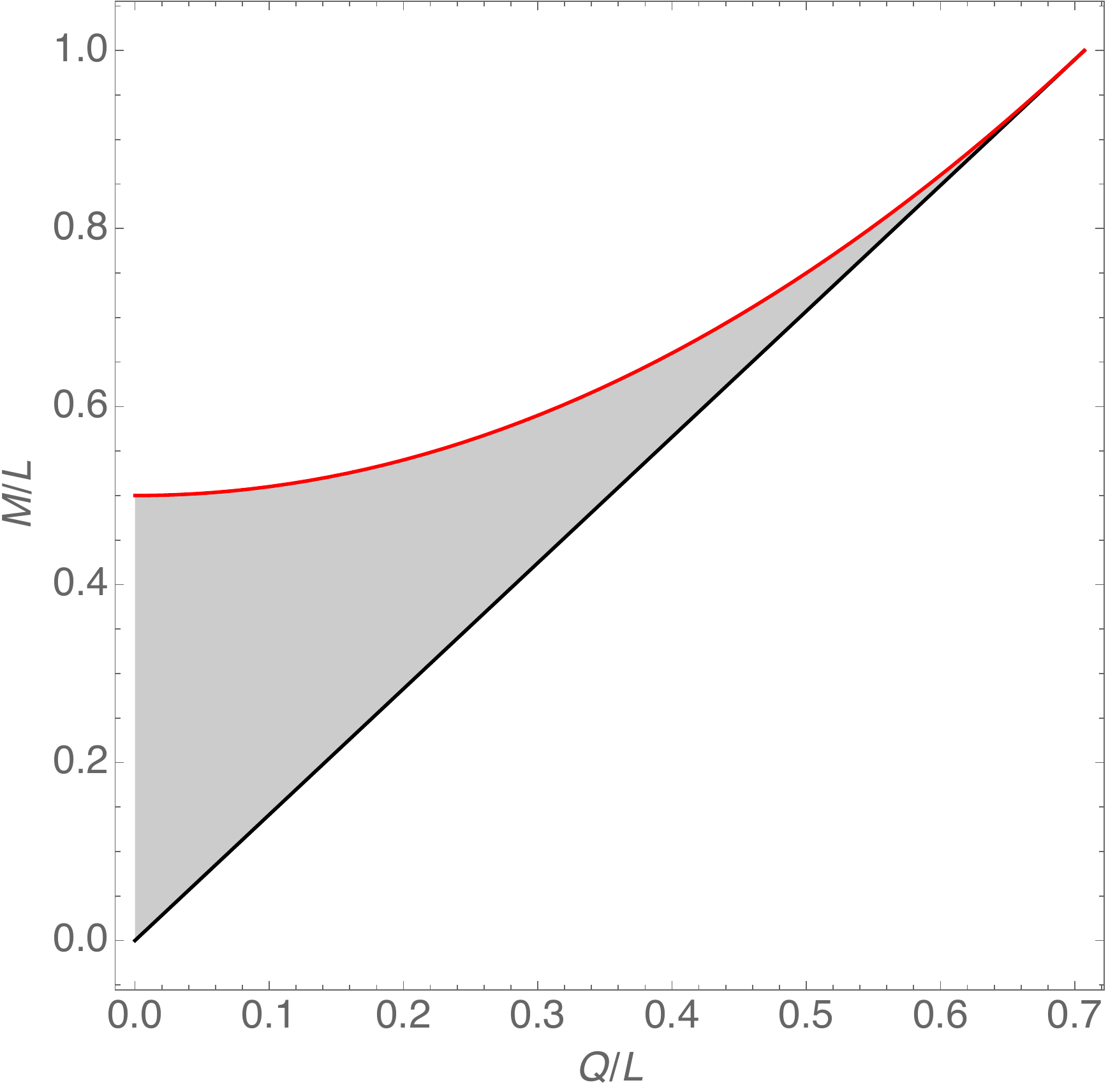}
}
\caption{Region of existence of RN BH inside the box in a phase diagram that plots the  adimensional ADM mass $M/L$ $vs$ the adimensional ADM electric charge $Q/L$. The lower black line corresponds to extremal RN BHs, and non-extremal RN BHs exist in all the region above this line. The upper red line describes RN BHs that have horizon radius equal to the radius of the box, $r_+=L$. Thus, the RN BHs that can fit inside the box are those in the shaded area limited by the two lines.}
\label{fig:charge_mass}
\end{figure}

The boxed RN BH system has the scaling symmetry ($x=\cos\theta$ and $\varphi$ are the polar and azimuthal angles of the $S^2$):
\begin{align}\label{1scalingsym}
\begin{split}
\{t,r,x,\varphi\}\to\{\lambda_1 t,\lambda_1 r,x,\varphi\} , &\qquad \{f,g,A,\varphi\}\to \{f,g,A,\varphi\}, \\
\{q,L, r_+,m\}&\to \left\{\frac{q}{\lambda_1},\lambda_1 L, \lambda_1 r_+,\frac{m}{\lambda_1}\right\}
\end{split}
\end{align}
which rescales the line element and the gauge field 1-form as $\mathrm{d}s^2\to \lambda_1^2 \mathrm{d}s^2$ and $A \mathrm{d}t \to \lambda_1 \,A \mathrm{d}t$ but leaves the action (equations of motion) invariant. We can use this scaling symmetry to work with dimensionless coordinates and measure thermodynamic quantities in units of $L$ (effectively this sets $L\equiv 1$),
\begin{equation}\label{adimTR}
T\equiv \frac t L\,,\qquad R\equiv\frac r L\,; \qquad R_+\equiv\frac{r_+}{L}\,, \qquad e\equiv q L\,, \qquad m_\phi \equiv m L \,.
\end{equation}
The box is now at $R=1$ and the condition that the BH is inside the box constrains the adimensional horizon radius to be $R_+<1$. In terms of the adimensional ADM mass and electric charge, RN BHs with $R_+<1$ and $0\leq \mu\leq \sqrt 2$  are those in the shaded region of  Fig. \ref{fig:charge_mass}.
 The (lower) black line corresponds to extremality ($\mu=\sqrt 2$) and the upper red line has $R_+=1$. 

To solve the Klein-Gordon equation \eqref{KG} as a  well-posed boundary value problem we must specify the boundary conditions that the scalar field must obey. 
We want  the scalar field to vanish at the box location $\Sigma$ and outside it, $\phi(R\geq 1)=0$. This implies we impose the outer boundary condition
\begin{equation}\label{BCbox} 
\phi \big|_{R=1}=0\,, 
\end{equation}
and we define the amplitude of the scalar field to be given by the value of its normal derivative at the box, $\phi^{\prime\:in} \big|_{R=1^-}\equiv \varepsilon$. 

At the inner boundary, $R=R_+$, the scalar field must be regular in Eddington-Finkelstein coordinates (which are appropriate to extend the background solution across the horizon). The scalar equation  \eqref{KG} is a second order PDE so there are two integration constants when we do a Taylor expansion about $R=R_+$. Regularity then requires that one of these constants (proportional to a divergent contribution) vanishes  and we are left with the arbitrary constant $\phi_0$ (say), to be determined by the equation of motion, such that the scalar field behaves as
\begin{equation}\label{BChorizon}
\phi\big|_{R=R_+}=\phi_0+\mathcal O((R-R_+)^2),
\end{equation}
In one of our discussions we will also be interested in the zero horizon radius limit, $R_+\to 0$ of the boxed RN BH. In this case, the system reduces to a scalar field perturbation of Minkowski spacetime inside a box. The inner boundary is then the origin, and regularity at this boundary requires that the scalar field behaves as $\phi\big|_{R=0}=\phi_0+\mathcal O(R)$, where the integration constant $\phi_0$ is determined once  \eqref{KG} subject to \eqref{BCbox} is solved. 

\section{Superradiant and near-horizon instabilities of a RN bomb}\label{sec:RNinstabilityHeuristics}

It is well-known that a charged scalar field perturbation in a boxed RN BH drives the system superradiantly unstable. However, one of our main results is the observation that this boxed RN system is also unstable to a second instability, namely the near-horizon scalar condensation instability (first identified in AdS BHs in the context of the holographic superconductor programme). Typically, these two instabilities co-exist {\ie} they are  entangled. However, we will show that in special limits of the parameter space they disentangle since one of the instabilities is suppressed. Thus, these special limits reveal their different nature. Each of these two limits will give analytical expressions for the onset of the instabilities. In the next section we will solve the Klein-Gordon equation numerically and find that the numerical solution for the instability onset indeed approaches these analytical results. 

In this section, we start by revisiting the origin and main properties of the superradiant instability (subsection \ref{subsec:Normal modes}). Then, in subsection \ref{subsec:nearinstability}, we will explain why  boxed RN BHs should also be unstable to the  near-horizon  scalar condensation instability

\subsection{Normal modes of flat spacetime with a box and superradiance in a boxed RN}\label{subsec:Normal modes}

It is enlightening to start our discussion with the case where we have a charged scalar field perturbation confined inside a box $\Sigma$ placed at $R=1$ (\ie $r=L$) in Minkowski space with a constant Maxwell field.  Naturally, this confinement condition, $\phi(R\geq 1)=0$, and regularity at the origin quantize the frequency spectrum of the scalar perturbations.
 
The electrovacuum background is time independent and axisymmetric. Therefore, 
we can Fourier expand our massive scalar field perturbation along the Killing directions $T$ and $\varphi$. This introduces the adimensional frequency  $\Omega=\omega L$ and the azimuthal quantum number $m_{\varphi}$. Moreover, in a linear mode analysis, we assume a separation {\it ansatz} for the radial and polar angle dependence. Altogether one has  
\begin{equation}\label{SeparationAnsatz}
\phi(T,R,x,\varphi)=e^{-i \,\Omega \,T}e^{i \,m_{\varphi} \varphi}P_{\ell}^{m_{\varphi}}(x)\,\psi(R)\,,
\end{equation}
where $Y_{\ell m_{\varphi}}=e^{i\,m_{\varphi}\varphi}P_\ell^{m_{\varphi}}(x)$ is a representation of the spherical harmonics, with $P_\ell^{m_{\varphi}}(x)$ being the associated Legendre polynomials ($\ell$ essentially gives the number of zeros of the eigenfunction along the polar direction). 

Using this ansatz, we can solve the Klein-Gordon equation \eqref{KG} in Minkowski spacetime with gauge field $A=\mu\, \mathrm{d}T$. The most general solution is
\begin{equation}
\psi(R)=\frac{\left[(\Omega+e\mu)^2-m_\phi^2\right]^{-\frac{1}{4}}}{\sqrt{R}}\left[ \beta_1 J_{\ell+\frac 1 2}\left(R \sqrt{(\Omega+e\mu)^2-m_\phi^2}\right)+\beta_2 Y_{\ell+\frac 1 2}\left(R \sqrt{(\Omega+e\mu)^2-m_\phi^2}\right)\right]\!\!,
\end{equation}
where $J_\nu(z)$ and $Y_\nu(z)$ are the Bessel functions of the first and second kind, respectively, and $\beta_1,\beta_2$ are arbitrary constants. 
We now have to impose the boundary conditions. Regularity of the solution at the origin, $R=0$, requires that we set $\beta_2=0$ to avoid a divergence of the type $R^{-\ell-1}$. On the other hand, at the location of the box, $R=1$, we want the scalar field to vanish; see \eqref{BCbox}. For $\nu \in \mathbb{R}$, the Bessel function $J_\nu(z)$ has an infinite number of simple zeros that we denote by $j_{\nu,n}$, \ie $n\in \mathbb{N}$ identifies the particular zero we look at. The boundary condition $\psi(R=1)=0$ then quantizes the frequency $\Omega$ in terms of the scalar field charge $e$ and black hole potential $\mu$. Altogether, the Klein-Gordon solution that obeys the required boundary conditions is 
\begin{equation}\label{leadingScalar}
\psi(R)=\beta_1 \frac{\left[(\Omega_{\ell,n}+e\mu)^2-m_\phi^2\right]^{-\frac{1}{4}}}{\sqrt{R}} \,J_{\ell+\frac 1 2}\left(R \sqrt{(\Omega_{\ell,n}+e\mu)^2-m_\phi^2}\,\right),
\end{equation}
with the frequency spectrum
\begin{align}\label{frequencies}
\Omega_{\ell,n}=\sqrt{j^2_{\ell+\frac{1}{2},n} +m_\phi^2}-e \mu\,,\quad n=1,2,3,\cdots \,.
\end{align}
Without loss of generality in what concerns the main physical results, onwards we restrict our attention to the spherically symmetric case, $\ell=0$, and we consider the solution with lowest frequency (\ie with lowest energy) which amounts to consider the first simple zero \ie $\ n=1$.\footnote{One has $j_{\frac 1 2,n}=n\pi$. In Fig. \ref{fig:eonsetharmonics} we will also display, as green diamonds at $R_+=0$, the information relative to $\ell=1$ and $\ell=2$ modes with $n=1$ and $m_\phi=0$.} Additionally, we will consider a massless scalar field, $m_\phi=0$. With these choices \eqref{frequencies} reads \footnote{\label{foot:groundstate}Note that we can use a $U(1)$ gauge transformation, $ \phi=|\phi|e^{i\varphi} \to |\phi|e^{i(\varphi +e \,\chi)}, \: A_t \to A_t +\nabla_t \chi$ to move to a frame where the frequency vanishes by choosing the chemical potential to be $\mu=\frac{1}{e}\sqrt{j^2_{\ell+\frac{1}{2},n} +m_\phi^2}$. For $\ell=0$ and $n=1$ this reads $\mu=\frac{\pi}{e}$.}
\begin{align}\label{eq:chemicalPotential}
\Omega_{0,1}=j_{\frac{1}{2},1} -e \mu=\pi-e \mu\,.
\end{align}
Condition \eqref{eq:chemicalPotential} gives the lowest frequency for a scalar field that can fit inside a box with radius $R=1$ in Minkowski spacetime. If this box is placed instead in a RN black hole background, the frequency spectrum will change. However, if the black hole is small, $R_+\ll 1$, the frequency can be approximated by a series expansion in $R_+$ with the leading order term being \eqref{eq:chemicalPotential}.
Borrowing results from a scalar field in AdS  \cite{Basu:2010uz,Dias:2016pma} we expect that, at higher orders in the $R_+$ expansion, the frequency correction  receives an imaginary contribution. From the results of \cite{Basu:2010uz,Dias:2016pma}, we further expect this imaginary contribution to be proportional to $\Omega_1$. Therefore, much like in the AdS case, the system should be unstable to superradiance for $\Omega_1<0$ (\ie  for $e\geq \pi/\mu$).
  In particular, for an extremal RN BH the chemical potential is $\mu=\sqrt 2$. It then follows that a small extremal RN BH should be unstable to superradiance for scalar field charges that obey the bound
\begin{equation}\label{eq:superradiantbound}
e\geq \frac{\pi}{\sqrt 2}\sim 2.221.
\end{equation}
This bound can be expanded in a series expansion in $R_+$ and away from extremality. It is enlightening to illustrate explicitly this expansion. In particular, the next-to-leading order correction of this expansion will later be useful to interpret our numerical results (see green dashed line in Fig. \ref{fig:eonset_Rp_chp}).

We expand the scalar field and the frequency as
\begin{equation}\label{expansionScalar}
\phi^{(\cal R)}(R)=\sum_{k\geq 0}R_+^{k}\phi_{k}^{(\cal R)}(R),\qquad \Omega=\sum_{k\geq 0}R_+^{k}\Omega_{k}\,,
\end{equation}
where the superscript $^{(\cal R)}$ indicates whether we are considering the near-region ($^{{(\cal R)}=near}$) or far-region ($^{{(\cal R)}=far}$) to be discussed next.
The leading order solution is given by \eqref{leadingScalar}-\eqref{eq:chemicalPotential} with $\Omega_0\equiv \Omega_{0,1}$ (again we consider the ground state solution with $n=1$). Unfortunately, the higher order solutions are harder to get because we cannot solve analytically the expanded equation of motion \eqref{KG}. Therefore, to be able to solve analytically the equations of motion we resort to a matched asymptotic expansion, similar to the one done in a similar context in AdS backgrounds in
\cite{Basu:2010uz,Bhattacharyya:2010yg,Dias:2011at,Dias:2011tj,Stotyn:2011ns}. 
Essentially, we divide the region inside the box into two sub-regions, namely the {\it near} region  $r_+\leq r\ll L$ ($^{near}$) and the {\it far} region where $r_+\ll r< L$  ($^{far}$). Considering small black holes with $r_+/L \ll 1$, these near and far regions have an overlapping zone, $r_+ \ll r\ll L$. In this overlapping region, we match the set of independent parameters that are generated by solving the perturbative equations of motion in the near and far regions. 
We have to be careful with the fact that in the near region, $R_+\leq R\ll 1$, Taylor expansion terms in $R_+ \ll 1$ can be of similar order as $R$. This is closely connected with the fact that the far region solution breaks down when  $R/R_+\sim \mathcal{O}(1)$. This indicates that in the near region analysis we should first introduce  a new radial, $y$, and time, $\tau$, coordinates as  
\begin{align}\label{eq:neartransform}
 y=\frac{R}{R_+}, \qquad \tau= \frac{T}{R_+}.
\end{align} 
Now, the near region corresponds to $1 \leq y \ll R_+^{-1}$. If we further require that $R_+\ll 1$ (as is necessarily the case in our perturbative expansion) one concludes that the near region corresponds to $R_+\ll 1\leq y$. In particular, Taylor expansions in $R_+ \ll 1$ can now be safely done since the radial coordinate $y$ and the horizon radius $R_+$ have a large hierarchy of scales.\footnote{A key step for the success of the matching expansion procedure is that a factor of $R_+$ (one of the expansion parameters) is absorbed in the new coordinates \eqref{eq:neartransform}.}
Further physical insight is gained if we rewrite the RN solution \eqref{eq:RNbackground} in the new coordinate system \eqref{eq:neartransform}: 
\begin{align}
&\mathrm d s^2=R_+^2\left(-f(y)\mathrm d \tau^2+\frac{\mathrm d y^2}{f(y)}+y^2\mathrm d \Omega_{2}^2\right),\qquad A_{\tau}(y)=R_+\;A_{T}(y)=R_+\;\mu\left(1-\frac{1}{y}\right), \nonumber \\
& \hspace{1cm}f(y)=1-\frac{2+\mu^2}{2 y}+\frac{\mu^2}{2 y^2},
\end{align}
The explicit factor of $R_+\ll 1$ in $A_{\tau}(y)$ shows that in the near region the electric field is weak. Thus, at leading order, the gauge field is suppressed in the equations of motion and the system is to be seen as a small perturbation around the neutral solution.

In these conditions, we want to solve the Klein-Gordon equation \eqref{KG} in the RN background, subject to the boundary condition \eqref{BCbox} at the box location $R=1$ and to regularity at the horizon, $R=R_+$. To impose smoothness of the perturbations at the horizon boundary it is a good idea to work with the ingoing Eddington-Finkelstein coordinate
\begin{equation}\label{EddFink}
v=T+\int \frac{1}{f(R)}\mathrm d R.
\end{equation}

We illustrate this procedure at leading order ($k=0$), \ie we recover \eqref{eq:chemicalPotential} this time using the matching expansion procedure. 
The most general solutions in the near and far regions are, respectively,
\begin{align}
\begin{split}
&\phi^{near}(y)=\sigma_1+\sigma_2\,\frac{\log (y-1)-\log \left(2 y-\mu ^2\right)}{2-\mu ^2},\\
&\phi^{far}(R)=\eta_1\, \frac{ e^{-i e \mu  R}}{R}+\eta_2\, \frac{e^{2\, i\, \Omega_0 R}}{\Omega_0+e \mu}\frac{e^{i e \mu  R}}{R}\,,
\end{split}
\end{align}
for  arbitrary integration constants $\{\sigma_1,\sigma_2\}$ and $\{\eta_1,\eta_2\}$.  Regularity of the near region solution at  $y=1$, requires that we eliminate the logarithmic  divergence $\log(y-1)$ by setting $\sigma_2=0$.
Requiring the  boundary condition \eqref{BCbox} at the box, $\phi|_{R=1}=0$, fixes $\eta_1=-\eta_2 \exp \left[2i(\Omega_0+e \mu)\right]/(\Omega_0+e \mu)$. Moreover, we fix our perturbation expansion parameter $\varepsilon$ to be such that  $\phi^{\prime} \big|_{R=1}\equiv \varepsilon$ (at all orders) which implies $\eta_2=-\frac{i}{2}\exp\left[-i(2\Omega_0+e \mu)\right]\varepsilon$. Next, we match the far and near regions. A small $R\ll 1$ Taylor expansion of the far region yields a $1/R$ term. We have to eliminate this contribution that is not present  in $\phi^{near}=\sigma_1$. This fixes the frequency to be $\Omega_0=\pi-e \mu$. Matching the ${\cal O}(1)$ near and far region contributions then fixes $\sigma_1=\varepsilon \,e^{i \,e\mu}$. Altogether, after imposing boundary and matching conditions one has:  
\begin{align}
\begin{split}
&\phi^{near}(R)=\varepsilon\, e^{i \,e\mu},\qquad \phi^{far}(R)=\varepsilon\,\frac{i e^{i \,e \mu (1-R)}}{2 \pi  R}\left(1-e^{2 i \pi  R}\right)\,;\\
&  \Omega_0=\pi-e \mu\,.
\end{split}
\end{align}
As it should, this value for  $\Omega_0 \equiv  \Omega_{0,1}$ agrees with the value 
\eqref{eq:chemicalPotential} we obtained solving exactly the leading order system.
This is the lowest frequency that can fit inside a box in Minkowski space.

The same matching asymptotic procedure with similar steps can be applied at next order, $\mathcal O(R_+)$. The expressions are now long and not very illuminating, so we immediately give the final value for the frequency correction $\Omega_1$:
\begin{align}
\Omega_1=\left(\mu e-\frac{1}{2} \pi  \left(\mu^2+2\right)\right) \Big[\gamma -\text{Ci}(2 \pi )+\log (2 \pi ) \Big]+\frac{1}{4} \pi  \left(\mu^2+2\right),
\end{align}
where $\gamma\sim 0.577216$ is Euler's constant and $\text{Ci}(x)=-\int_x^{\infty}\frac{\cos z}{z}\mathrm d z$ is the cosine integral function.
We could now proceed with the expansion to higher order $\mathcal O\left(R_+^k\right)$, to find the successive frequency corrections $\Omega_k$ for $k\geq 2$. However, we do not do it here since we will solve the Klein-Gordon equation numerically in the next section to get the full frequency. 
For our purposes, the correction $\Omega\simeq \Omega_0+ \Omega_1 R_+$ is already enough. In the gauge we work, the onset of superradiance occurs when $\Omega=0$. We can then use the condition $ \Omega_0+ \Omega_1 R_+ +\mathcal{O}\left(R_+^k\right)= 0$ to find the electric charge of the scalar field, $e_{onset}$ as a function of the horizon radius and chemical potential, above which the system is unstable to superradiance. This gives:
\begin{equation}\label{eq:onsetSuperradiance}
e_{onset}=\frac{\pi}{\mu}-R_+\,\frac{\pi}{2\mu}\left[ \mu^2 \Big(\gamma+\log (2 \pi )-\rm{Ci}(2 \pi )-\frac{1}{2} \Big)-1\right]+\mathcal{O}\left(R_+^2\right).
\end{equation}

In Section \ref{sec:QNM} we will solve  numerically the linearised Klein-Gordon equation to find the linear instabilities of a RN BH in a box. With a numerical analysis we will confirm that, in the small horizon radius limit and for RN black holes near extremality, the system indeed becomes unstable if \eqref{eq:superradiantbound} and \eqref{eq:onsetSuperradiance} are obeyed.
In particular, in Fig. \ref{fig:eonset_Rp_chp}, the green dashed straight line is given by \eqref{eq:onsetSuperradiance}  when we set the chemical potential to its extreme value, $\mu=\sqrt 2$. 

\subsection{The origin of the near-horizon instability of RN black holes in a box}\label{subsec:nearinstability}
It was first found in the context of holographic superconductors \cite{Gubser:2008px,Hartnoll:2008kx,Hartnoll:2008vx} in planar AdS spacetimes that charged near extremal black holes can be unstable to near horizon scalar condensation. It was then understood that this instability is a generic feature of near extremal black holes in asymptotically AdS spacetimes \cite{Dias:2010ma}. Since  a box in asymptotically flat spacetimes has confining/reflecting boundary conditions similar to those of asymptotically AdS spacetimes, it is natural to ask whether this near horizon instability is also present in the setup we are considering.

To address this question in generic conditions, consider a scalar field with mass $m$ and charge $q$ (we find it convenient to restore the dimensionful quantities in this discussion). Stability conditions in flat spacetime require the mass $m$ to be non-negative, $m\geq 0$. We can follow the original analysis in asymptotically AdS spacetimes \cite{Dias:2010ma,Dias:2011tj,Dias:2016pma} to find the near horizon geometry of an extremal, asymptotically flat RN black hole.
Consider the RN metric \eqref{eq:RNbackground}, we introduce the new coordinates $\{\tau,\tilde{\rho}\}$,
\begin{align}\label{eq:NHcoordinates}
t=L_{AdS_2}^2 \,\frac{\tau}{\lambda},\qquad r=r_+ +\lambda \tilde{\rho},
\end{align}
and take the near horizon  limit $\lambda\to 0$. At leading order the metric takes the form
\begin{align}\label{eq:NHmetric}
&\mathrm{d}s^2=L_{AdS_2}^2\left( -\tilde{\rho}^2 \mathrm{d}\tau^2+\frac{\mathrm{d}\tilde{\rho}^2}{\tilde{\rho}^2} \right)+r_+^2\mathrm{d}\Omega_{(2)}^2, \nonumber \\
&A_{\mu}\mathrm{d}x^{\mu}=\alpha\, \tilde{\rho} \,\mathrm{d}t 
 \qquad \hbox{with}\:\:\:  L_{AdS_2}=r_+\:\:\: \hbox{and}\:\:\:  \alpha \equiv\sqrt{2}L_{AdS_2}.
\end{align}
Thus the near horizon geometry is the direct product of an AdS$_2$ geometry with an $S^2$.  This is still a solution of \eqref{action} \ie of Einstein-Maxwell theory\footnote{On the other hand the $AdS_2$ orbit space is a solution of 2-dimensional Einstein-AdS theory whose equation of motion, in the trace reversed form, reads $R_{ij}+L_{AdS_2}^{-2}g_{ij}=0$.}.

The Klein Gordon equation \eqref{KG} in a static background of the form \eqref{fieldansatz} can be written as :
\begin{align}
\partial_r \left( r^2 f \,\partial_r \psi \right)+\left( \frac{r^2}{f}\left(\omega+q A_t\right)^2 -\ell  (\ell +1) -m^2 r^2\right)\psi =0.
\end{align}  
To capture the time dependence information, the frequency $\tilde{\omega}$ in the near-horizon geometry is related to $\omega$ as:
\begin{align}
\omega=\tilde\omega \frac{\lambda}{L^2_{AdS_2}}.
\end{align}
which follows from the identification $e^{-i\omega t}\equiv e^{-i \tilde{\omega} \tau}$.

Taking the near horizon limit $\lambda\to 0$ of the Klein Gordon equation one gets
\begin{align}
\partial_{\tilde{\rho}}\Big(\tilde{\rho}^2\,\partial_{\tilde{\rho}} \psi(\tilde{\rho})\Big)+\left(\frac{\left(\tilde\omega+q \,\alpha \,\tilde{\rho} \right)^2}{\tilde{\rho}^2}-m^2 L_{AdS_2}^2-\ell(\ell+1)\right)\psi(\tilde{\rho})=0.
\end{align}
A Taylor expansion of this equation at large $\tilde{\rho}$ yields
\begin{align}
\psi{\bigl |}_{\tilde{\rho}\to\infty}\simeq a \,\tilde{\rho}^{\,-\widetilde{\Delta}_-}
+\cdots+ b \,\tilde{\rho}^{\,-\widetilde{\Delta}_+} +\cdots\,, \quad
 \hbox{with} \quad \widetilde{\Delta}_\pm=\frac{1}{2}\pm \frac{1}{2}\sqrt{1+4 m_{eff}^2L_{AdS_2}^2}\;,
\end{align}
with an effective mass for the scalar field given by
\begin{eqnarray}
\label{nearmass}
m_{eff}^2 L_{AdS_2}^2&\equiv&  m^2 L_{AdS_2}^2+\ell(\ell +1) -q^2 \,\alpha^2 \nonumber\\
 &=&  m^2 L_{AdS_2}^2+\ell(\ell +1)-2 q^2 r_+^2.
\end{eqnarray}
A scalar field in AdS$_2$ is unstable if it violates the AdS$_2$ Breitenl\"ohner-Freedman \cite{Breitenlohner:1982jf} bound, \ie if its mass \eqref{nearmass} violates the bound
\begin{align}
m_{eff}^2 L_{AdS_2}^2\geq -\frac 1 4\,.
\end{align}
Therefore, it is natural to expect that extremal RN BHs confined in a box might be unstable to scalar condensation when their scalar field charge is large enough, namely when
\begin{align}\label{eq:NHcharge}
q\geq \frac{\sqrt{(2\ell+1)^2+4 m^2 L_{AdS_2}^2}}{2 \sqrt{2} r_+}=\frac{\sqrt{(2\ell+1)^2+4 m^2 r_+^2}}{2 \sqrt{2} r_+}.
\end{align}

Now we reintroduce the dimensionless quantities $R_+=r_+/L$, $e=q L$ and $m_\phi=m L$. For large black holes inside the box, a Taylor expansion of \eqref{eq:NHcharge} about  $R_+\sim 1$ yields
\begin{align}\label{eq:NHinstabilityLargeRp}
\begin{split}
&e\geq \frac{\sqrt{(2\ell+1)^2+4 m_\phi^2}}{2\sqrt 2}+\frac{(2\ell+1)^2}{2 \sqrt{2} \sqrt{1+4( m_\phi^2+l(l+1))}}(1-R_+)+\mathcal O\Big((1-R_+)^2\Big).
\end{split}
\end{align}
On the other hand, for small black holes \eqref{eq:NHcharge} has the expansion,
\begin{equation}\label{eq:NHinstabilitySmallRp}
e\geq \frac{2\ell+1}{2 \sqrt{2}}\, \frac{1}{R_+}+\mathcal O(R_+^1).
\end{equation}
Thus, for small black holes $R_+\ll 1$, we see that the near horizon instability is suppressed. However, near horizon condensation is certainly present for large black holes (when compared with the box radius). In particular in the limit $R_+\to 1$, and for massless scalar fields that is our main focus, the instability is present for charges
\begin{align}\label{eq:NHbound}
e\gtrsim \frac {2\ell+1} {2\sqrt 2}.
\end{align}

This is in contrast with the superradiant instability that is present for small black holes and suppressed for large black holes. The near horizon procedure strictly applies to extremal black holes. However, continuity suggests that this instability is present away from extremality.

In the next section we will compute numerically the timescales of linear instabilities of a RN BH in a box. We will find that as the system approaches extremality and $R_+\to 1$, the instability indeed has  a near horizon condensation origin since its onset will be given by \eqref{eq:NHcharge}  (see the right panel of Fig. \ref{fig:eonset_Rp_chp} and Fig. \ref{fig:eonsetharmonics}).

\section{Timescale for superradiant and near-horizon instabilities}\label{sec:QNM}

\begin{figure}[t]
\centerline{
\includegraphics[width=.48\textwidth]{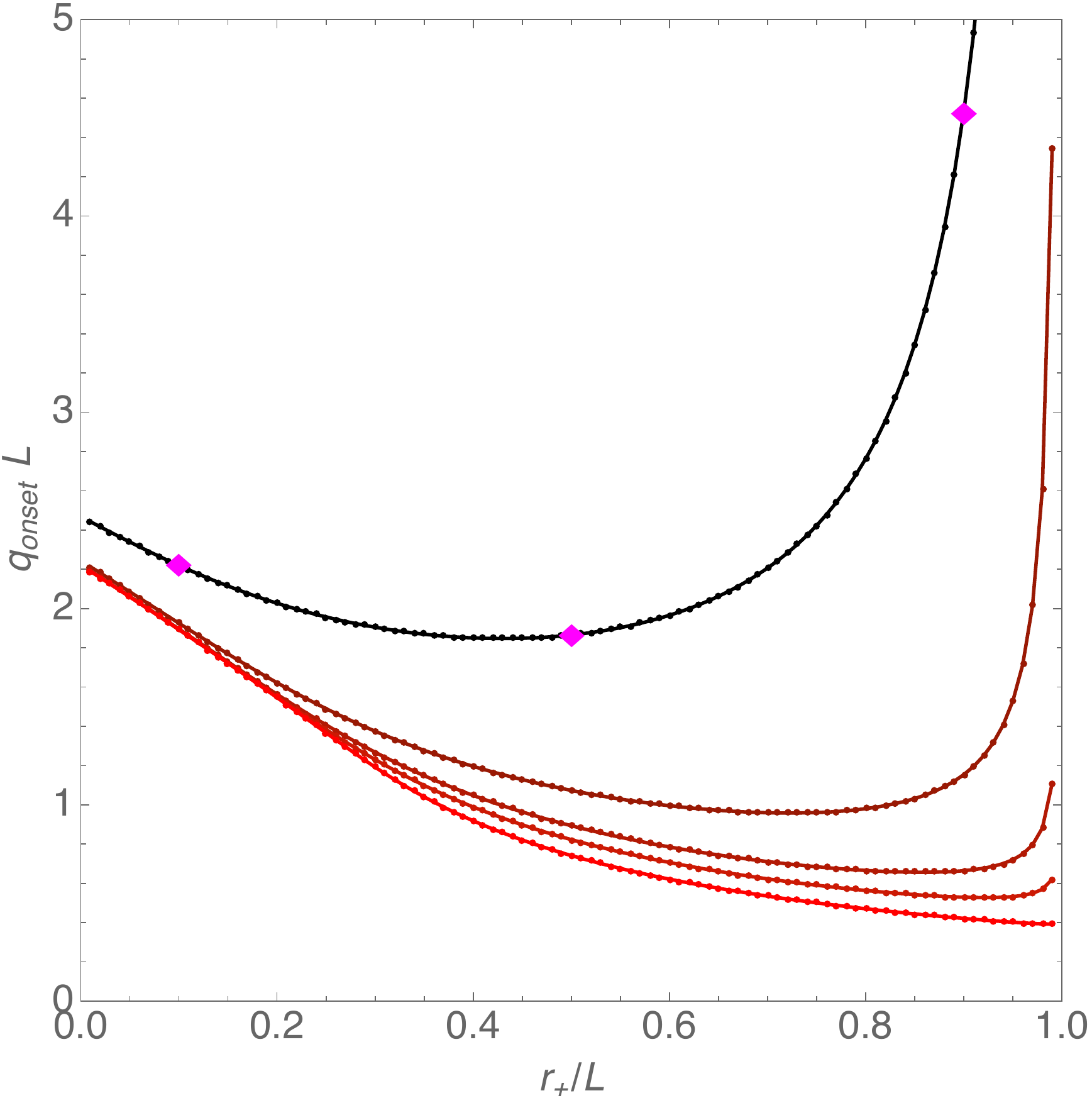}
\hspace{0.5cm}
\includegraphics[width=.48\textwidth]{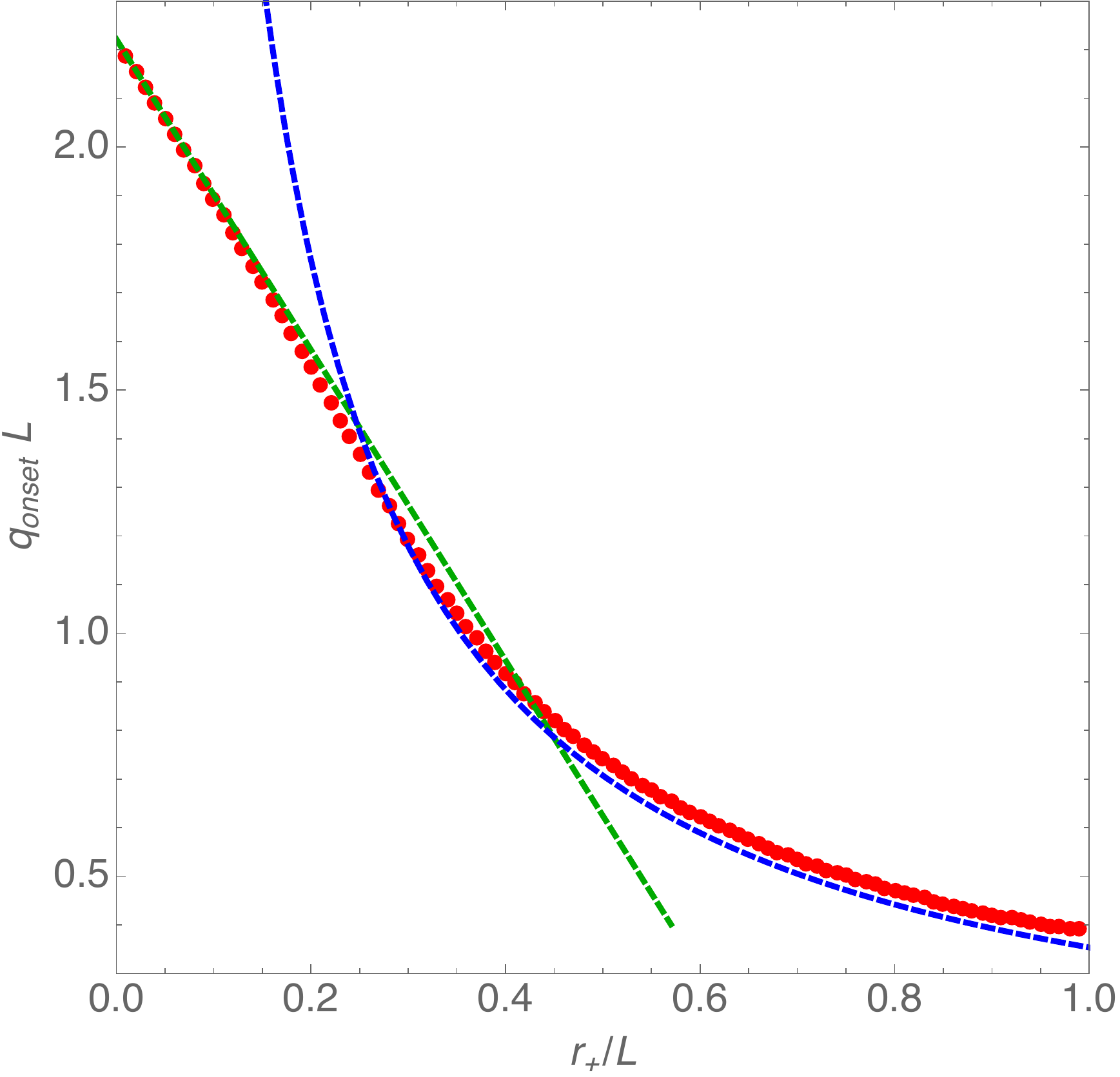}
}
\caption{{\bf Left panel:} Electric charge $e_{onset}=L q_{onset}$ of the scalar field that signals the onset of instabilities for RN BHs in a box as a function of the horizon radius $R_+=r_+/L$, and for 5 constant values of the chemical potential $\mu=\sqrt 2 (1-10^{-n})$ with $n=1,2,3,4,8$ from top (black) to bottom (red). Above these lines, RN BH are unstable as the imaginary part of the frequency $\Omega$ is positive. The magenta diamonds pinpoint three particular RN BHs:  in the right panel of Fig. \ref{fig:eonset_MQ} we explicitly show that ${\rm Im}\,\Omega$ changes sign precisely at $e=e_{onset}$ identified in this figure.
{\bf Right panel:} Similar to left panel but this time we just plot the onset curve (dots) with $\mu=\sqrt 2 (1-10^{-8})$ (closest to extremality), and we add the analytical curves  \eqref{eq:onsetSuperradiance} (dashed green straight line with negative slope) and \eqref{eq:NHcharge} (dashed blue curve).
}
\label{fig:eonset_Rp_chp}
\end{figure}

In this section we compute numerically the frequencies of a massless scalar field confined inside a box in a RN background. In particular, we find the scalar field charge above which linear mode instabilities appear.  This problem was already addressed in previous literature for particular values of the parameters involved in the problem. However, here we want to complement these available results with  sharper statements about the region in phase space where the solutions are unstable. For example, for a given scalar field mass\footnote{For concreteness, in the presentation of our results, we will fix the scalar field mass to zero, but the computation can be repeated for different masses $m$. In all cases, the onset curve $e_{onset}(R_+,\mu;m)$ for the instability approaches the analytical limiting values \eqref{eq:superradiantbound} and \eqref{eq:NHbound}.} we want to identify the critical scalar field charge $e_{onset}(R_+,\mu)$ above which instabilities are present. More importantly, we want to point out that RN BHs are linearly unstable not only to the superradiant instability but also to the near-horizon scalar condensation instability. Previous studies missed the existence of the latter instability. The two instabilities are usually highly entangled (which justifies why previous numerical studies missed identifying the existence of the near-horizon instability) but near extremality they disentangle. In this regime, we can explicitly check that the  numerical onset curve $e_{onset}(R_+,\mu)$ approaches the analytical expression for superradiance \eqref{eq:superradiantbound}  and the analytical expression \eqref{eq:NHbound} for the near-horizon instability   in the appropriate limits, namely for small and large horizon radius $R_+$, respectively (see the right panel of Fig. \ref{fig:eonset_Rp_chp}). 

Consider the  Klein-Gordon equation \eqref{KG} in the RN spacetime \eqref{fieldansatz}-\eqref{eq:RNbackground}. Like in the Minkowski background, we can assume the separation ansatz \eqref{SeparationAnsatz} for the scalar field, $\phi(T,R,x,\varphi)=e^{-i \,\Omega \,T}e^{i \,m_{\varphi} \varphi}P_{\ell}^{m_{\varphi}}(x)\,\psi(R)$. Due to spherical symmetry of the background, the azimuthal number $m_{\varphi}$ does not contribute to the radial equation. Recall that $\Omega=\omega L$ is the dimensionless frequency\footnote{\label{foot:Gauge}Note that $\omega$ is the frequency in the gauge of \eqref{eq:RNbackground} whereby  the gauge potential $A_t\big|_{R_+}=0$ and $A_t\big|_{R\to \infty}=\mu$. Alternatively, it is also common to choose the gauge $A_t=-\mu R_+/R$ in which case $A_t$ vanishes asymptotically and the associated scalar field frequency $\widehat{\omega}$ is related to $\omega$ by $\omega=\widehat{\omega}- q \mu$. Later, we will find that the system is unstable when ${\rm Re} \,\omega<0$ which corresponds to ${\rm Re} \,\widehat{\omega}< q \mu$.}. In these conditions, solving the Klein-Gordon equation boils down to solving the radial ODE,
\begin{equation}\label{eq:KGeigenmodes}
\partial_R \left( R^2 f \,\partial_R \psi \right)+\left( \frac{R^2}{f}\left(\Omega+\mathit{e} A_t\right)^2 -\ell  (\ell +1) \right)\psi =0
\end{equation}
with $f(R)$ and $A_t(R)$ being the RN background functions defined in  \eqref{eq:RNbackground}. We are mostly interested in the most unstable mode which happens to be the one with $\ell=0$ that preserves the spherical symmetry. We will confirm this is indeed the case. 

This equation has to be solved subject to appropriate boundary conditions. 
A Taylor expansion of the Klein-Gordon equation about the horizon yields the two linearly independent solutions:
\begin{equation}
\psi(R)\big|_{R\to R_+}\sim \beta_{in}(R-R_+)^{-\frac{i  \Omega}{4\pi T_H L}}+\beta_{out}(R-R_+)^{\frac{i  \Omega}{4\pi T_H L}}+\cdots.
\end{equation}
Regularity of the perturbation in Eddington-Finkelstein coordinates \eqref{EddFink} requires that we choose the boundary condition $\beta_{out}=0$ (this effectively discards outgoing modes and keeps the ingoing waves). 

On the other hand, a Taylor expansion around the location $R=1$ of the box yields:  
\begin{equation}
\psi(R)\big|_{R\to 1}\sim \psi_{0} +\psi_{1} (R-1) +\cdots.
\end{equation}
We want the scalar field to vanish at the location of the box so we impose the Dirichlet boundary condition $ \psi_{0}=0$.

In order to solve numerically equation \eqref{eq:KGeigenmodes} we introduce the new function $p(R)$ as
\begin{align}\label{eq:changefunction}
\psi(R)=(R-R_+)^{-\frac{i  \Omega}{4\pi T_H L}}(1-R)\, p(R).
\end{align}
In addition, we find  it convenient (it yields a Neumann boundary condition at the horizon) to work with the new radial variable
\begin{align}\label{eq:changevariables}
y= \left( \frac{R-R_+}{1-R_+}\right)^{\frac{1}{2}}.
\end{align}
Coordinate $y$ ranges between $y=0$ (\ie $R=R_+$) and $y=1$ (\ie $R=1$). In terms of the new function $p(y)$, the boundary conditions read simply as
\begin{align}\label{BCsNum}
p'(0)=0\,,\qquad p'(1)+ \left(\frac{\left(\mu ^2+2\right) R_+-4}{\mu ^2 R_+-2}-\frac{4 i R_+ \Omega}{2-\mu ^2}\right)p(1)=0.
\end{align}

The ODE  \eqref{eq:KGeigenmodes}  and the boundary conditions \eqref{BCsNum} yield a quadratic eigenvalue problem in the frequency $\Omega$, for a given $e$. Interestingly, if we we want to find the onset of an instability (\ie which has $\Omega=0$), it is also a quadratic eigenvalue problem in the electric scalar field charge $e$. We will explore both perspectives in our analysis. 

For the numerical discretization scheme of $y\in [0,1]$ we use pseudo-spectral methods with a Chebyshev grid  (see details \eg in \cite{Dias:2015nua}). 
We solve for the eigenvalue and associated eigenfunction using one of two methods. In one method, we make use of the fact that the equation at hand is a quadratic eigenvalue problem, which can thus be solved using \emph{Mathematica}'s built-in routine \emph{Eigensystem} (see \eg \cite{Dias:2010eu,Dias:2015nua} for details). The second method is based on an application of the Newton-Raphson root-finding algorithm, and is detailed in \cite{Cardoso:2013pza,Dias:2015nua}. The advantage of the first method is that it gives many modes simultaneously (\ie many radial overtones), allowing for an easy identification of the spectra. The second method computes a single mode family at a time, and can be used  when we have an educated guess for a seed sufficiently close to the true answer. Yet, this method is computationally more efficient, and can be used to push the numerics to extreme regions of the parameter space. 
Details of these numerical methods and discretization scheme can be found in the review \cite{Dias:2015nua}. Numerical convergence tests are presented in Appendix \ref{App:Numerics}.

We can now discuss the results. The system is unstable when the imaginary part of the frequency $\Omega=\omega L$ is positive.  An indication of the onset of the instability is given when we encounter what are called the zero-modes which correspond, in our gauge choice, to $\Omega=0$, \ie ${\rm Re}\,\Omega=0={\rm Im}\,\Omega$ (see footnote \ref{foot:Gauge}). In fact, as discussed above, we can set $\Omega=0$ and solve \eqref{eq:KGeigenmodes} and \eqref{BCsNum} as an eigenvalue problem for the scalar field charge $e\equiv e_{onset}$. This determines a surface expanded by the chemical potential and the horizon radius. Instead of displaying a 3-dimensional plot, we find more enlightening to pick different values of the chemical potential $\mu$ and plot the dependence of the onset charge $e_{onset}$ as a function of the horizon radius $R_+$. This plot is shown in the left panel of Fig. \ref{fig:eonset_Rp_chp}. 
The curves with the dots describe the onset charge  $e_{onset}$ as a function of the horizon radius for a given choice of chemical potential $\mu$. More concretely, we consider 5 values for  $\mu$. Namely, we set $\mu=\sqrt 2 (1-10^{-n})$ and from the top (black curve) to the bottom (red curve) the 5 curves have $n=1,2,3,4,8$.

\begin{figure}[t]
\centerline{
\includegraphics[width=.48\textwidth]{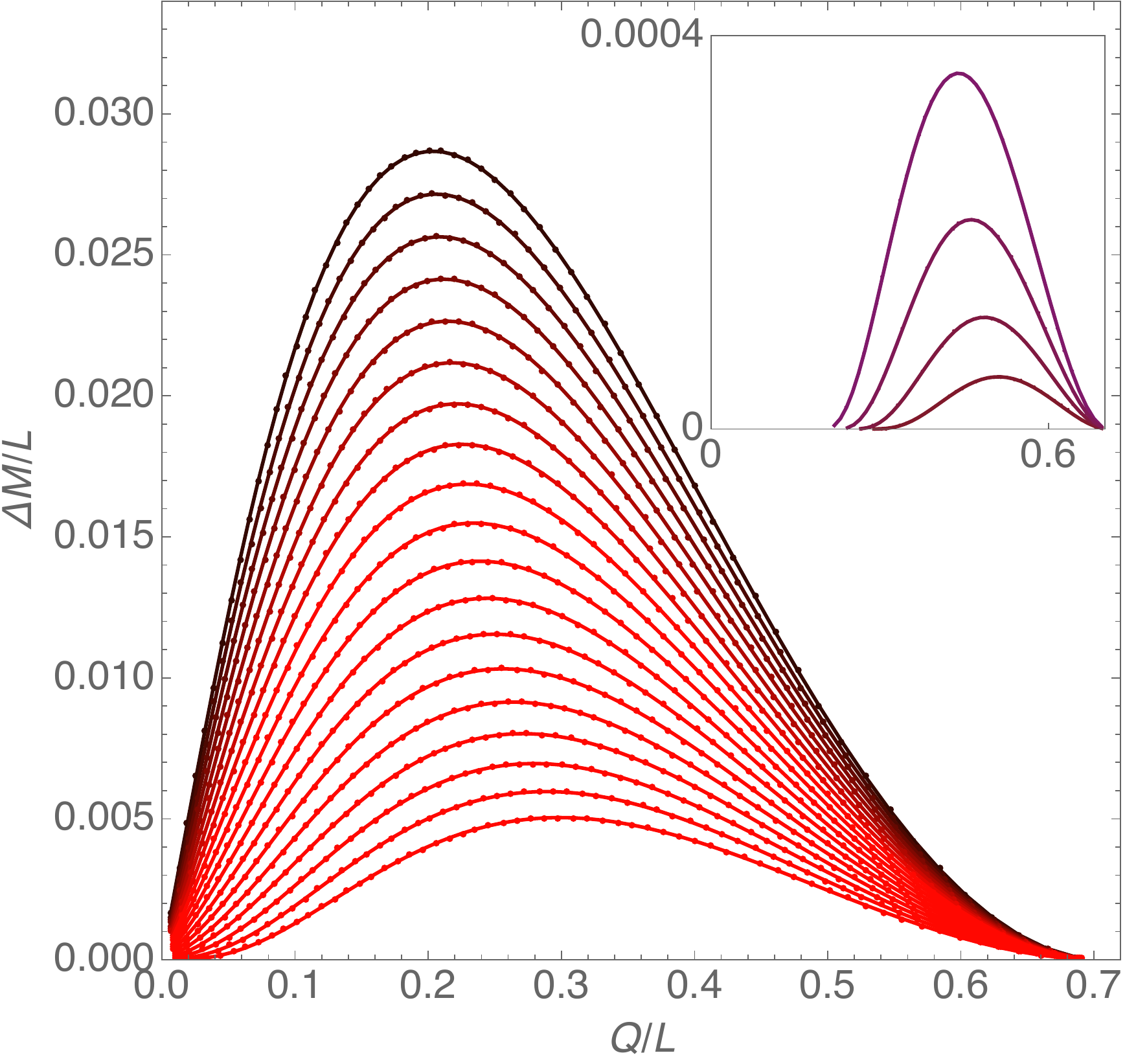}
\hspace{0.5cm}
\includegraphics[width=.48\textwidth]{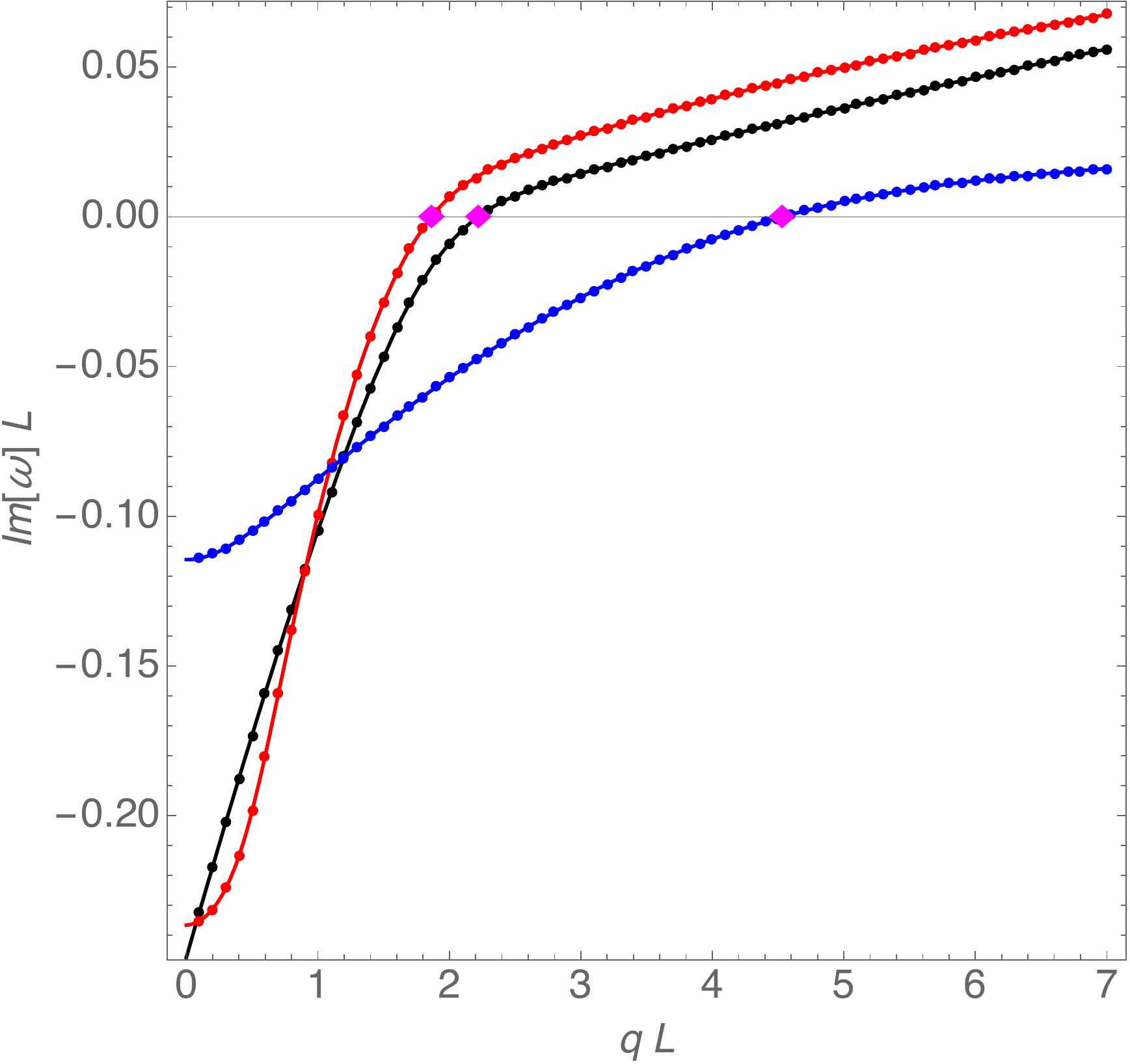}
}
\caption{{\bf Left panel:} Onset curves in a phase diagram that plots the RN electric charge $Q/L$ as a function of the mass difference $\Delta M/L$ w.r.t. the extremal RN with same $Q/L$. The several curves have different constant scalar field charge: $e_{onset}=4$ (upper curve) all the way down (in steps of $-0.1$) to $e_{onset}= 2.2$ (bottom curve).  The curves in the inset plot correspond to scalar field charges $e_{onset}=1.3$ (top) down to $e_{onset}=1$, in steps of $-0.1$.  {\bf Right panel:} Imaginary part of the dimensionless frequency ${\rm Im}\,(\omega L)$ as a function of the dimensionless scalar field charge $e=q L$ at constant chemical potential $\ \mu=\sqrt 2(1-10^{-1})$ and constant horizon radius: $R_+=0.1$ (black), $R_+=0.5$ (red) and  $R_+=0.9$ (blue). The magenta diamonds correspond precisely to the onset charge $e=e_{onset}$ already identified in Fig. \eqref{fig:eonset_Rp_chp}.}
\label{fig:eonset_MQ}
\end{figure}

To further interpret the left panel of Fig. \ref{fig:eonset_Rp_chp}, note that, previously in the literature (first for a scalar field in a Kerr box \cite{Cardoso:2004nk,Hod:2016rqd} and then for a charged scalar field in a RN box \cite{Herdeiro:2013pia,Hod:2016kpm}) it was found that, for a given scalar field charge, there is a minimum critical radius where the mirror can be placed to have an instability. If the mirror is placed below this critical radius there is no instability, that is to say, would-be unstable modes cannot fit inside the box. The superradiant condition allows a straightforward interpretation of this result: recalling footnote \ref{foot:Gauge}, in the gauge where the gauge potential vanishes asymptotically, superradiant modes have a maximum frequency, ${\rm Re} \,\widetilde{\omega}< q \mu$. Superradiant modes thus have a minimum wavelength and the mirror must be placed at a radius larger than this wavelength if we want to have unstable modes.  Now, in our study we use a scaling symmetry of the system to fix the mirror at the dimensionless radius $R=1$. It follows that the above conclusion restates as follows: for a given RN BH (i.e. for a given $\{R_+, \mu\}$) unstable modes are present if and only if the scalar field charge is above a critical value $e>e_{onset}(R_+,\mu)$. Fig. \ref{fig:eonset_Rp_chp} precisely displays these critical curves $e_{onset}(R_+,\mu)$ for selected values of $\mu$. These curves are exact (for $m_\phi=0$) and sharpen onset bounds given previously in \cite{Herdeiro:2013pia,Hod:2016kpm}. 

In the left panel of Fig. \ref{fig:eonset_Rp_chp},  the lowest numerical (red) curve  has $\mu=\sqrt{2}(1-10^{-8})$ i.e. it is extremely close to the extremal RN solution which has $\mu=\sqrt{2}$. This curve is particularly enlightening to understand the underlying physical nature of the unstable modes, and how the origin of the instability changes as we go from small to large radius black holes (in units of the mirror radius). For this discussion, in the  right panel of \ref{fig:eonset_Rp_chp} we take the bottom curve of the left panel with  $\mu=\sqrt 2 (1-10^{-8})$ and we add two analytical  curves.  The dashed green straight line with negative slope (starting at $R_+=0$ and $e_{onset}=\pi/\sqrt{2}\sim 2.22$) is the superradiant instability curve \eqref{eq:onsetSuperradiance},  which is valid for small $R_+$. On the other hand, the blue dashed curve (starting on the right at $R_+=1$ and $e_{onset}=\frac 1 {2\sqrt 2}\sim 0.354$) corresponds to the near-horizon instability curve \eqref{eq:NHcharge} with $m_\phi=0$, $e_{onset}=(2 \sqrt{2} R_+)^{-1}$.  We conclude that as extremality ($\mu=\sqrt 2$) is reached, the onset curve approaches the critical values predicted by the analytical expressions \eqref{eq:onsetSuperradiance} and \eqref{eq:NHcharge} (with $m_\phi=0$ and $\mu=\mu_{ext}=\sqrt{2}$).
 This is a check of our numerical results. More importantly,  it clearly demonstrates one of our main results: RN BHs in a box are not only afflicted by the superradiant instability (as studied in previous literature) but also by the near-horizon scalar condensation instability. For a generic RN BH these two instabilities are entangled but in the RN extremal limit they disentangle and their different nature unravels: the superradiant instability is present for small black holes but suppressed for large black holes while the near-horizon instability has the opposite behavior (it is supressed for small black holes as indicated by \eqref{eq:NHinstabilitySmallRp}).

\begin{figure}[t]
\centerline{
\includegraphics[width=.48\textwidth]{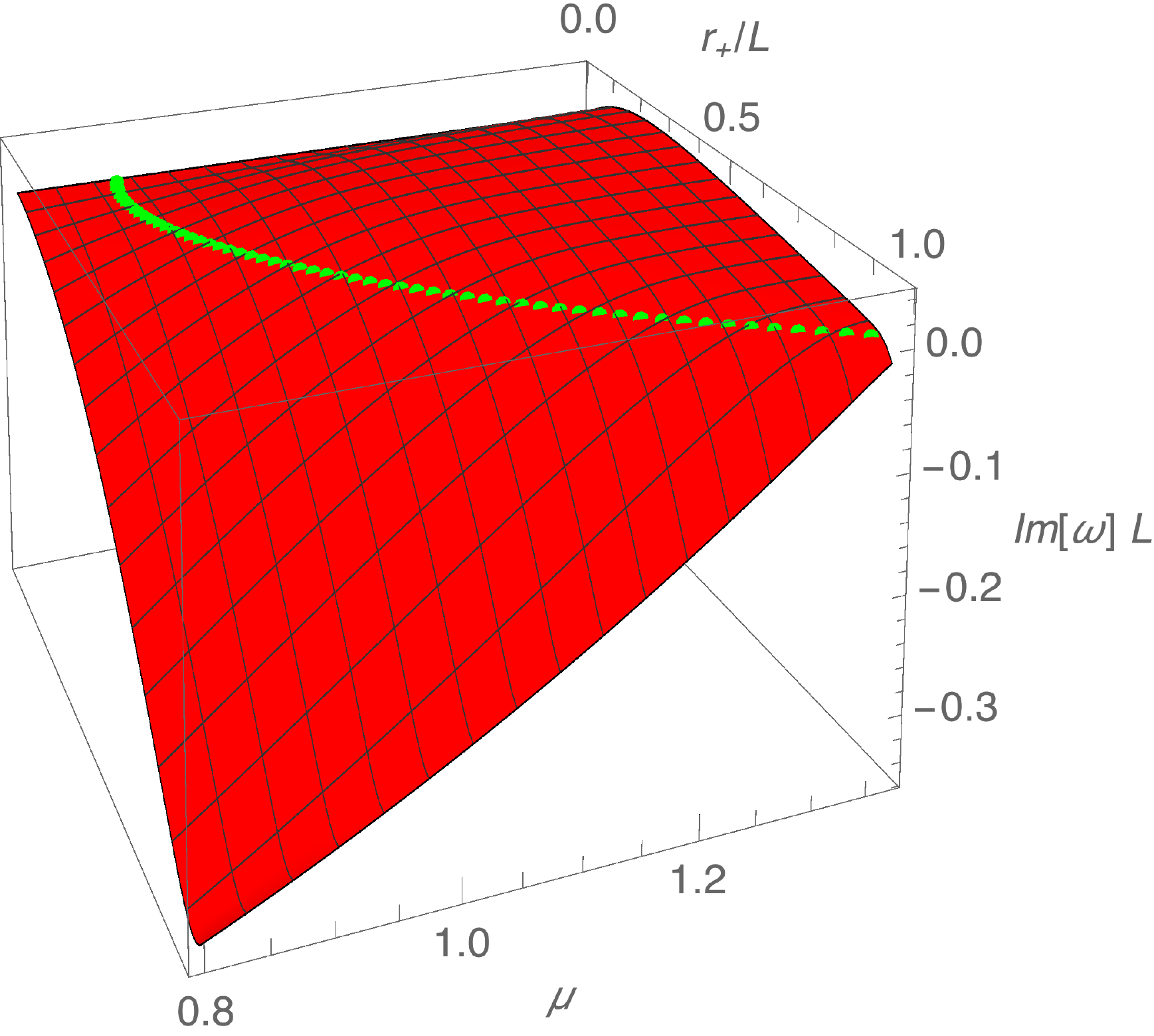}
\includegraphics[width=.48\textwidth]{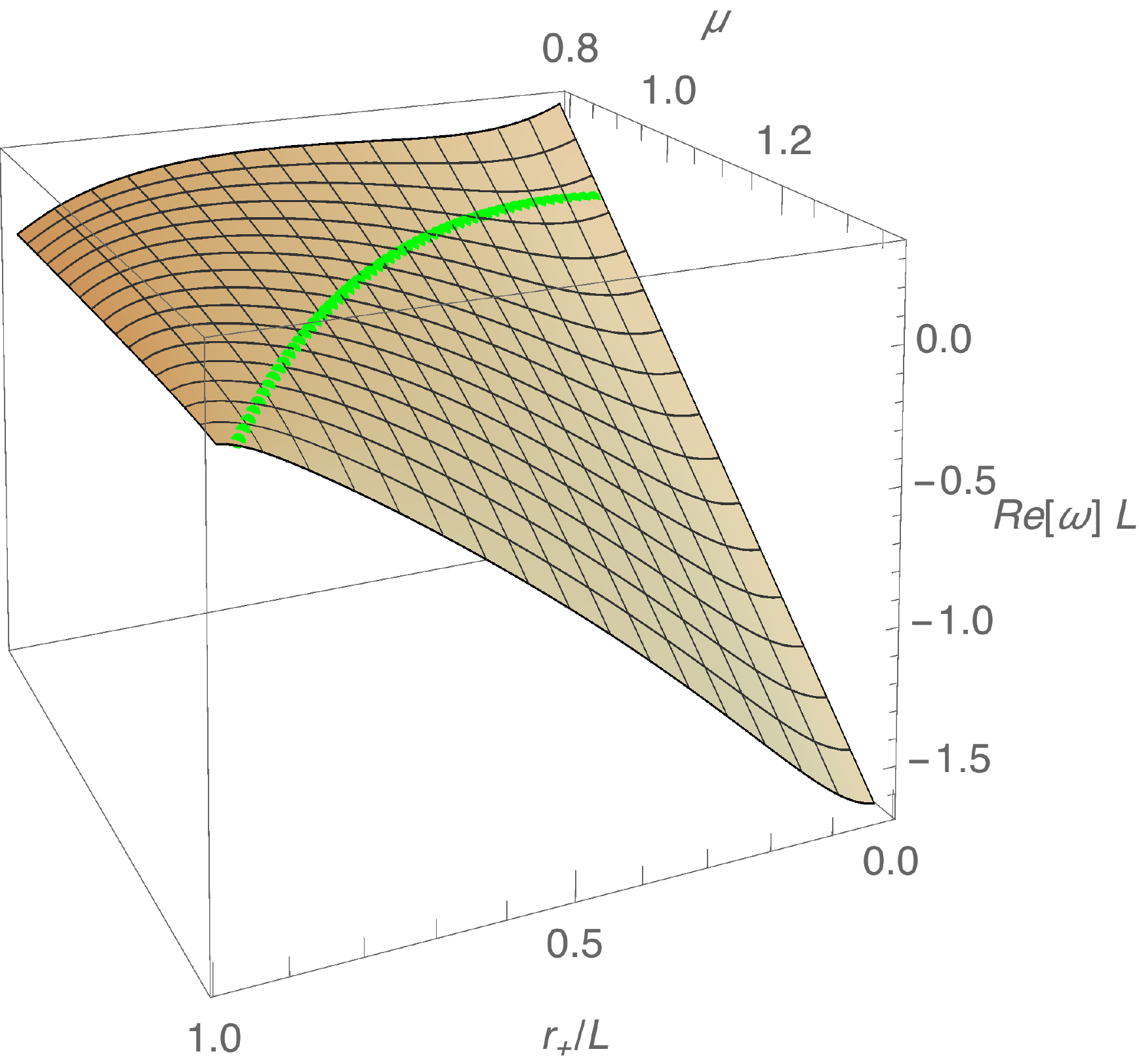}
}
\caption{The imaginary part ({\bf left panel})  and real part ({\bf right panel}) of the frequencies as a function of the chemical potential $\mu$ and the horizon radius $r_+$ for $e=q L=3.5$.  The dotted green curve  is the onset curve with ${\rm Re}(\omega L)=0$ and ${\rm Im}(\omega L)=0$.}
\label{fig:Re_Im_e}
\end{figure}

The information in Fig. \ref{fig:eonset_Rp_chp} is complemented by  Fig. \ref{fig:eonset_MQ}. In the left panel, on the horizontal axis we have the dimensionless ADM charge $Q/L$ of the RN BH. On the vertical axis, we have the mass difference $\Delta M/L=(M-M_{ext})/L$ between a RN BH and the extremal RN BH that has the same electric charge $Q/L$. So extremal RN BHs are represented by the horizontal line with $\Delta M=0$. We display several curves in the main left panel plot. Each curve corresponds to a fixed value for the onset charge $e_{onset}$ (\ie an horizontal line in Fig. \ref{fig:eonset_Rp_chp}): the top curve starts with $e_{onset}=4$ and then we go down in steps of $\delta e_{onset}=-0.1$ to the curve in the bottom that has $e_{onset}=2.2$. For each $e_{onset}$ curve, RN BHs below the onset curve are unstable. We see that, starting from $Q=0$, as $Q$ grows so does the region that is unstable: black holes with larger $M/L$ that are further away from extremality become unstable. However, for each $e_{onset}$ curve, $\Delta M/L$ attains a maximum around $Q/L\sim 0.2$. For larger $Q/L$,  $\Delta M/L$ starts decreasing monotonically and approaches zero as $Q/L$ approaches its maximum value $\sim 0.7$, which corresponds to $R_+\to 1$.

It is often stated that a RN in a Minkowski box resembles a RN in global  anti-de Sitter (AdS) spacetime: reflecting boundary conditions at the timelike boundary of AdS mean that AdS effectively behaves as a box with AdS radius $L$ (set by the cosmological constant $\Lambda=-3/L^2$). Although this statement is generically correct, there are also important differences between the two systems. Notably, the box in the RN BH is at a finite proper distance from the horizon, while in AdS the asymptotic timelike boundary is at an infinite proper distance. This key difference manifests in properties of the instabilities. This is particularly clear 
when we compare the behaviour displayed in Fig. \ref{fig:eonset_MQ} with its AdS counterpart. In AdS, all onset curves $e_{onset}$ are monotonic: larger $Q/L$ corresponds to larger $\Delta M/L$ (below which the system is unstable) and there is no bound for the electric charge $Q/M$ (neither for $M/L$); see \cite{Dias:2011tj} for the corresponding plot for RN-AdS$_5$ BHs. For small RN BHs inside a box, the system behaves similarly to the AdS one only for small $Q/L$: initially $\Delta M/L$ of the onset curve indeed increases as  $Q/L$ grows. However, unlike in AdS, here the $e_{onset}$ curve has a maximum $\Delta M/L$ and, for larger RN charge,  $Q/L$ itself also  has a maximum value where   $\Delta M/L\to 0$. So above $Q/L \sim 0.2$ (say), the box at finite proper distance induces an effect that differs from the behaviour in AdS. This behaviour can be understood from Fig. \ref{fig:charge_mass}: as we increase the charge $Q/L$ and the mass $M/L$, the region of existence of RN BHs inside a box becomes smaller and shrinks to zero as one approaches $Q/L\sim 0.7$. This is not the case in global AdS: unstable RN-AdS BHs exist for arbitrarly large charge and mass.

\begin{figure}[t]
\centerline{
\includegraphics[width=.48\textwidth]{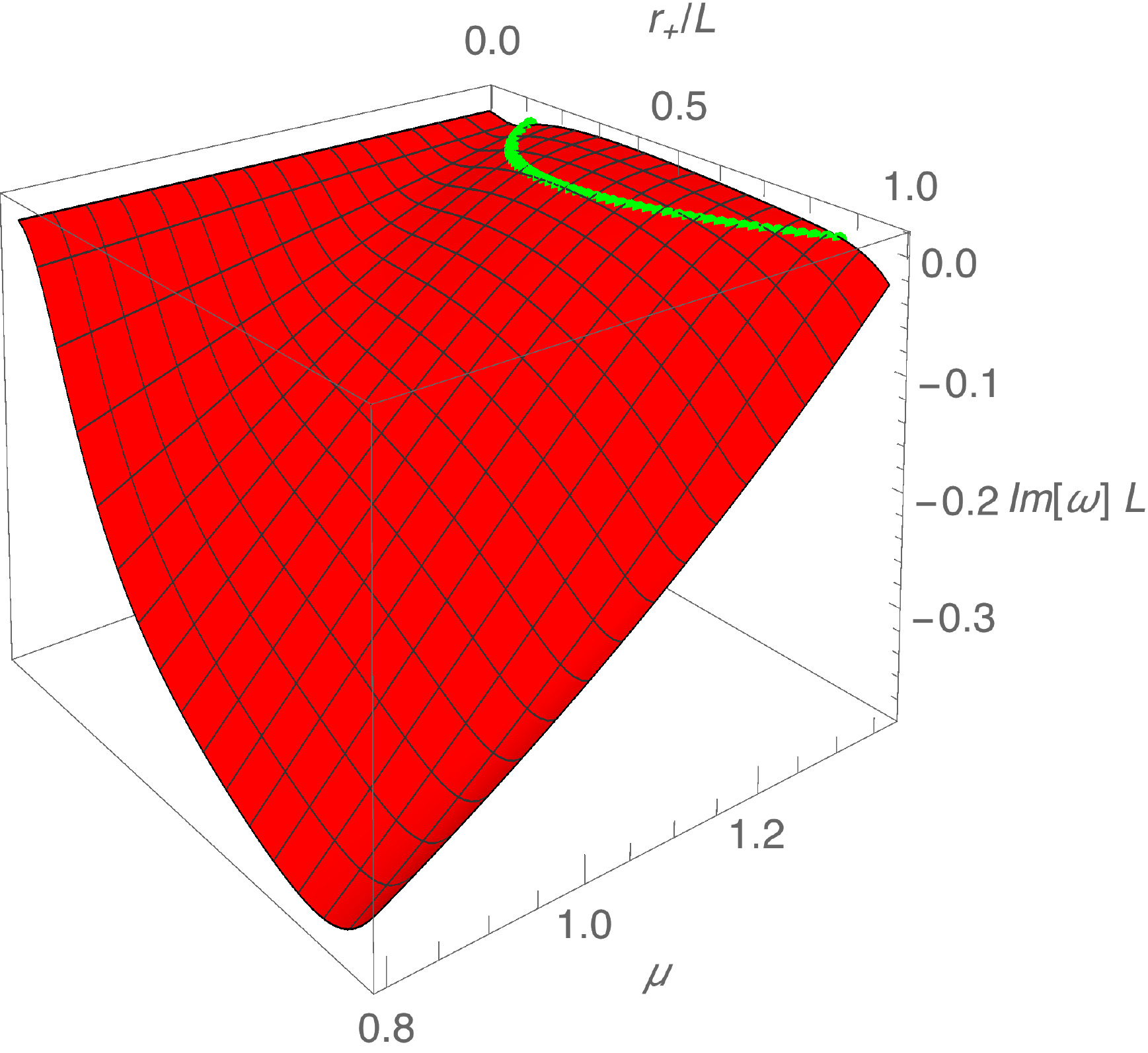}
\includegraphics[width=.48\textwidth]{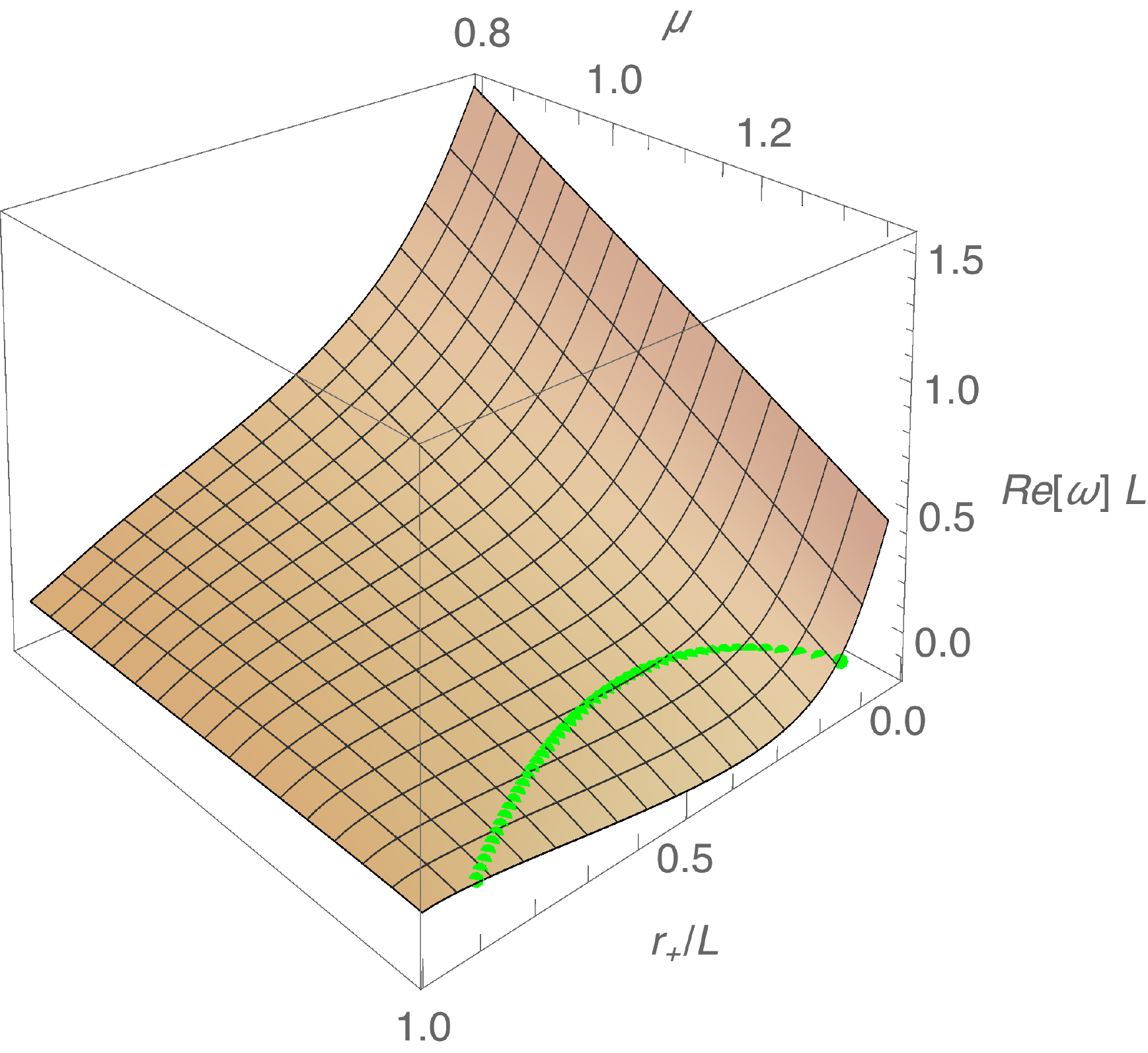}
}
\caption{Similar to Fig. \ref{fig:Re_Im_e} but this time for a scalar field charge of $e=1.9$. The dotted green curve  is the onset curve with ${\rm Re}(\omega L)=0$ and ${\rm Im}(\omega L)=0$.}
\label{fig:Re_Im_e1.9}
\end{figure}

Still in the left panel of Fig. \ref{fig:eonset_MQ}, in the inset plot we show similar curves for scalar field charges $e_{onset}=\{1.3,1.2,1.1,1\}$ which are all below  $\frac{\pi}{\sqrt 2}\sim 2.22$. We see that for $e \leq\frac{\pi}{\sqrt 2}$ the onset does not extend to all values of the charge $Q/L$ and mass $M/L$ since these $e$'s lie in the region $\frac{1}{2\sqrt 2}\leq e \leq\frac{\pi}{\sqrt 2}$ (see Fig. \ref{fig:eonset_Rp_chp}). Indeed, one can use the bottom curve in the Fig. \ref{fig:eonset_Rp_chp}, since $\mu=\sqrt 2 (1-10^{-8})$ is very near extremality, as a good estimative of the extremal curve. The curves in the inset plot of the right panel of Fig. \ref{fig:eonset_MQ} correspond to horizontal lines in Fig. \ref{fig:eonset_Rp_chp} with $\frac{1}{2\sqrt 2}\leq e \leq\frac{\pi}{\sqrt 2}$ that start at the extremal curve.

To check independently that the onset curves (surface) in Fig. \ref{fig:eonset_Rp_chp}  indeed signal the onset of the instability we have computed the quasinormal mode frequencies  $\Omega=\omega L$: above the onset surface the imaginary part of $\Omega$ becomes positive. For a simple check, we  chose specific values of the horizon radius $R_+$ and chemical potential $\mu$ and compute the imaginary part of the frequency ${\rm Im}\, \Omega$ as a function of the scalar field charge $e=q L$. We show three of these curves in the right panel of Fig. \ref{fig:eonset_MQ}. All these curves have $\ \mu=\sqrt 2(1-10^{-1})$ but different horizon radius, $R_+=\{0.1,0.5,0.9\}$. These particular RN BHs correspond to the magenta diamonds pinpointed in the upper curve of the left panel of Fig. \ref{fig:eonset_Rp_chp}. That is to say, at the value  $e=e_{onset}$ identified in the left plot of Fig. \ref{fig:eonset_Rp_chp} the imaginary part of the frequency becomes zero in Fig. \ref{fig:eonset_MQ} (magenta dots in each curve) which is a check to our numerics as these are two independent computations. 

\begin{figure}[t]
\centerline{
\includegraphics[width=.5\textwidth]{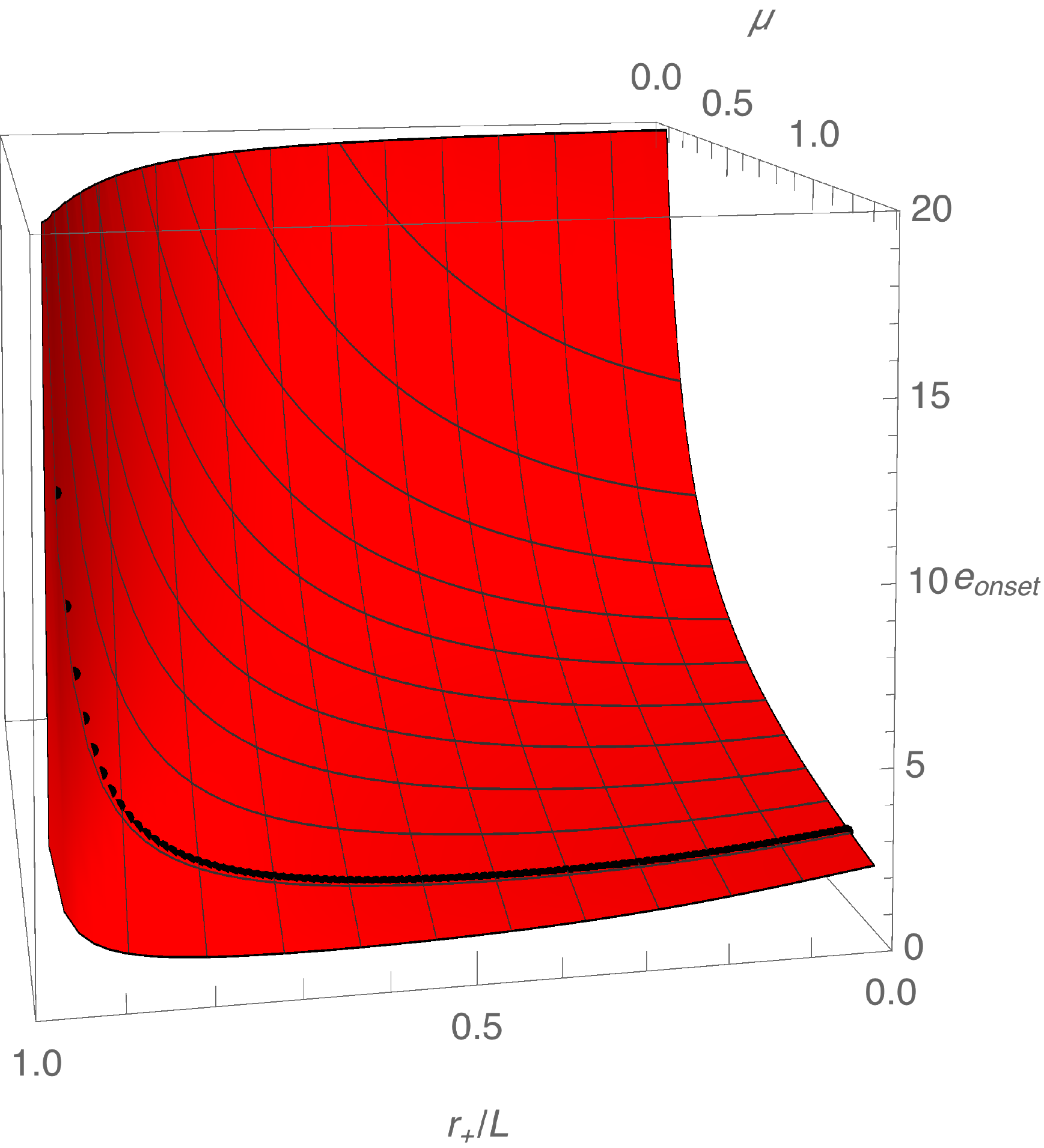}
}
\caption{Critical onset charge $e_{onset}$  as a function of the horizon radius $R_+$, and of the chemical potential $\mu$. For reference, the black dots represent the curve with $\mu=\sqrt 2 (1-10^{-1})$ already plotted in the top of the left panel of Fig.~\ref{fig:eonset_Rp_chp}.}
\label{fig:onset3d}
\end{figure}

For completeness, in Fig. \ref{fig:Re_Im_e} we fix the  scalar field charge $e=3.5$ and we show a complete 3D plot of both the imaginary and real parts of the dimensionless frequency $\Omega=\omega L$ as a function of the horizon radius $R_+$ and chemical potential $\mu$. The instability is present when ${\rm Im}\,\Omega>0$. As a non-trivial test of our computations we have checked that the instability onset curve for $e_{onset}=3.5$ in the left panel of Fig. \ref{fig:eonset_MQ} coincides with the onset curve in this 3D plot: dotted green curves with ${\rm Im}\,\Omega=0$ and ${\rm Re}\,\Omega=0$.  Values of $e$ above $\pi/\sqrt{2}\sim 2.221$ $-$ see \eqref{eq:onsetSuperradiance} and Fig. \ref{fig:eonset_Rp_chp}$-$ typically give plots qualitatively similar to those of Fig. \ref{fig:Re_Im_e}. Namely, the instability is present from $R_+=0$ all the way up to a maximum $R_+<1$, as long as we are sufficiently close to extremality.   

On the other hand, scalar field charges in the range $\pi/\sqrt{2} <e < 1/(2\sqrt{2})\sim 0.354$  $-$ see \eqref{eq:onsetSuperradiance}, \eqref{eq:NHbound} and Fig. \ref{fig:eonset_Rp_chp}$-$  give plots qualitatively similar to those in Fig. \ref{fig:Re_Im_e1.9}. In this figure, we fix $e=1.9$ and plot the imaginary and real parts of the frequency as a function of the horizon radius $R_+$ and chemical potential $\mu$. For this range of $e$, the instability is present only in a window of $R_+$ (so starting at $R_+\neq 0$), and when we are close to extremality.

Fig. \ref{fig:onset3d} complements Fig. \ref{fig:eonset_Rp_chp} since it shows the evolution of the onset charge $e_{onset}$ with $R_+$ for a larger range of the chemical potential $\mu$. Namely, for $\mu\leq \sqrt{2}\left(1-\frac{1}{61}\right)$ (for larger $\mu$ see Fig. \ref{fig:eonset_Rp_chp}). We see that as we move away from extremality and as the box  approaches the horizon, it becomes considerably more difficult to drive the system unstable. In the sense that a higher $e>e_{onset}$ is required to make the system unstable. 

So far all our data refers to modes with $\ell=0$. This is always the most unstable mode since it is the mode that: 1) requires a smaller $e_{onset}$, 2) is unstable for a wider range of black hole parameters $\{R_+,\mu\}$ (given a $e$), and 3) it is the mode that has the fastest growth rate, (smaller timescale $\tau=1/{\rm Im}(\omega L)$).  
 
\begin{figure}[t]
\centerline{
\includegraphics[width=.48\textwidth]{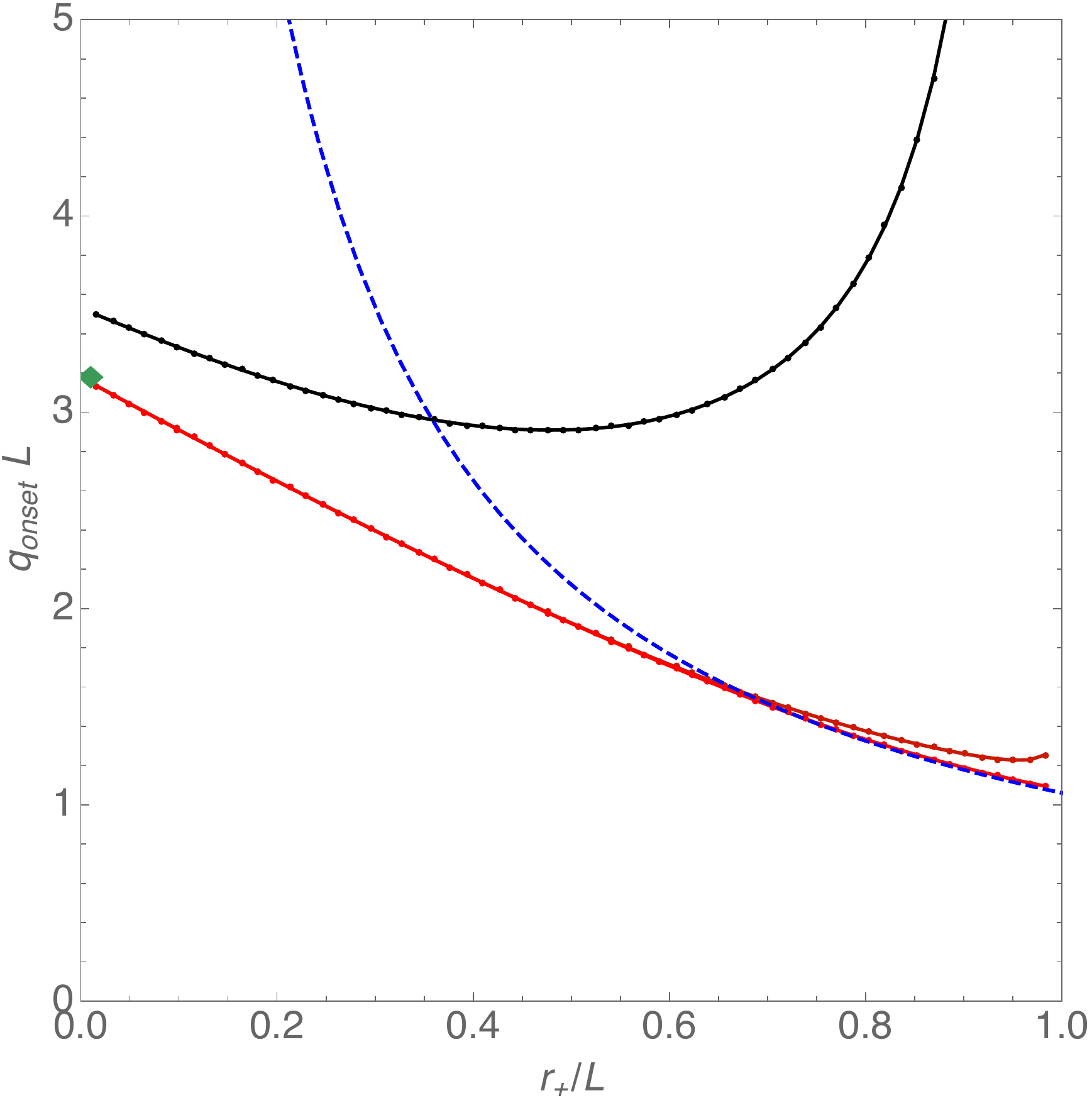}
\hspace{0.5cm}
\includegraphics[width=.48\textwidth]{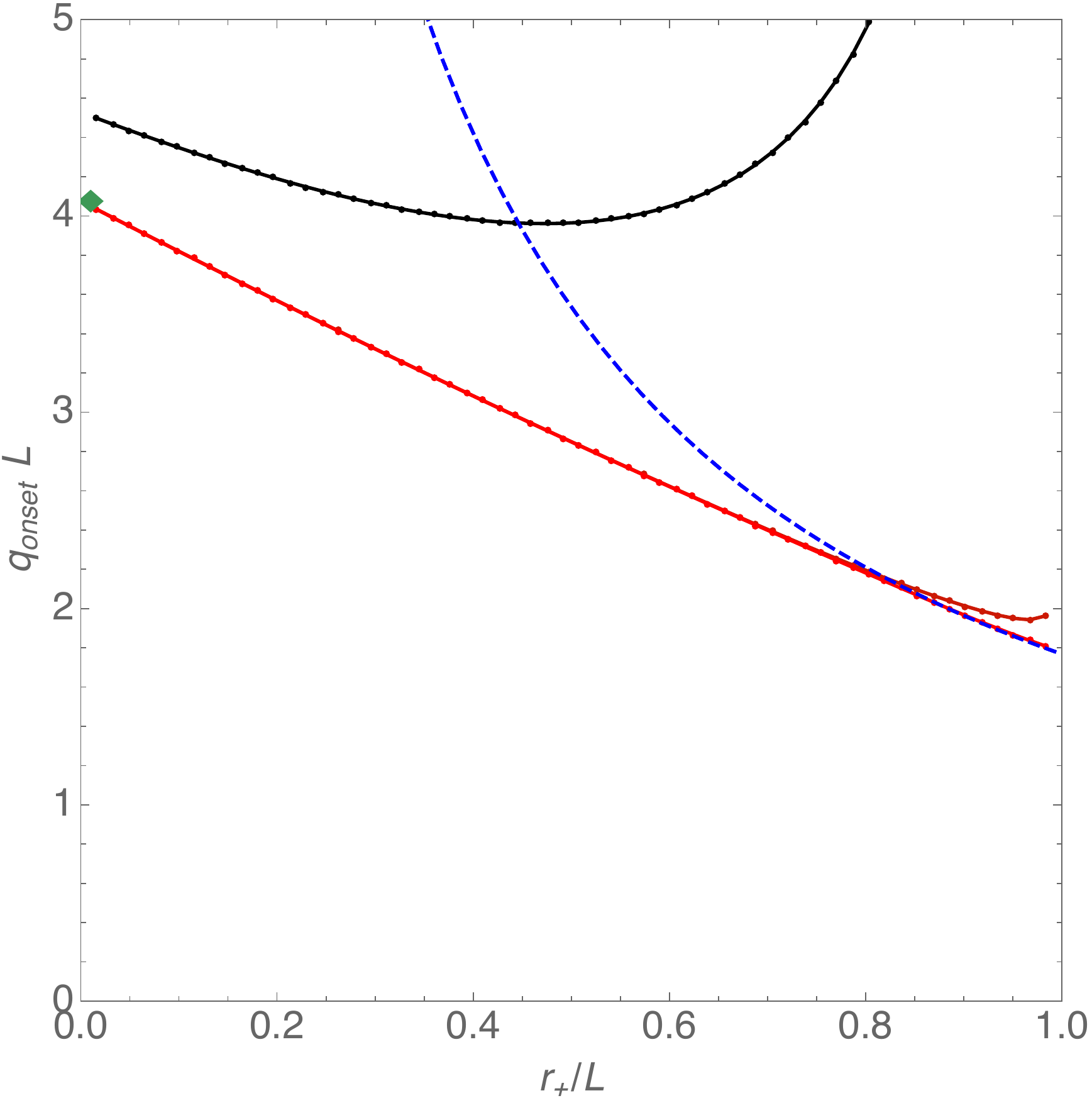}
}
\caption{Onset charge $e_{onset}$ as a function of $R_+$ for three chemical potentials, 
 $\mu=\sqrt 2 (1-10^{-n})$ with $n=1,4,8$ from top (black) to bottom (red). {\bf Left panel:} $\ell=1$ mode. The green diamond is at $\{R_+,e_{onset}\}=\{0,3.17732\}$ and the dashed blue curve is the near horizon prediction \eqref{eq:NHinstabilityLargeRp}.  {\bf Right panel:} $\ell=2$ mode. The green diamond is at $\{R_+,e_{onset}\}=\{0,4.07538\}$ and the dashed blue curve is the near horizon prediction \eqref{eq:NHinstabilityLargeRp}. 
}
\label{fig:eonsetharmonics}
\end{figure}

 Property 1) is illustrated in Fig. \ref{fig:eonsetharmonics} where we show the minimal onset charge $e_{onset}(R_+)$ (that is reached at extremality) for the $\ell=1$ (left panel) and $\ell=2$ (right panel) modes. This is to be compared with the similar plot for the $\ell=0$ mode shown in Fig. \ref{fig:eonset_Rp_chp}. Also shown in Fig. \ref{fig:eonsetharmonics} are the analytical superradiant (section \ref{subsec:Normal modes}) and near-horizon (\ref{subsec:nearinstability}) predictions. We conclude that they are also extremely sharp for higher $\ell$ modes. Refs. \cite{Herdeiro:2013pia,Hod:2016kpm} have given lower analytical bounds for $e_{onset}$; see in particular (44) of \cite{Hod:2016kpm}. They are indeed obeyed by our data but they are not sharp, in the sense that they can be well below (in some cases by factors of 2 and 10) the true onset charge. 
 
\begin{figure}[t]
\centerline{
\includegraphics[width=.48\textwidth]{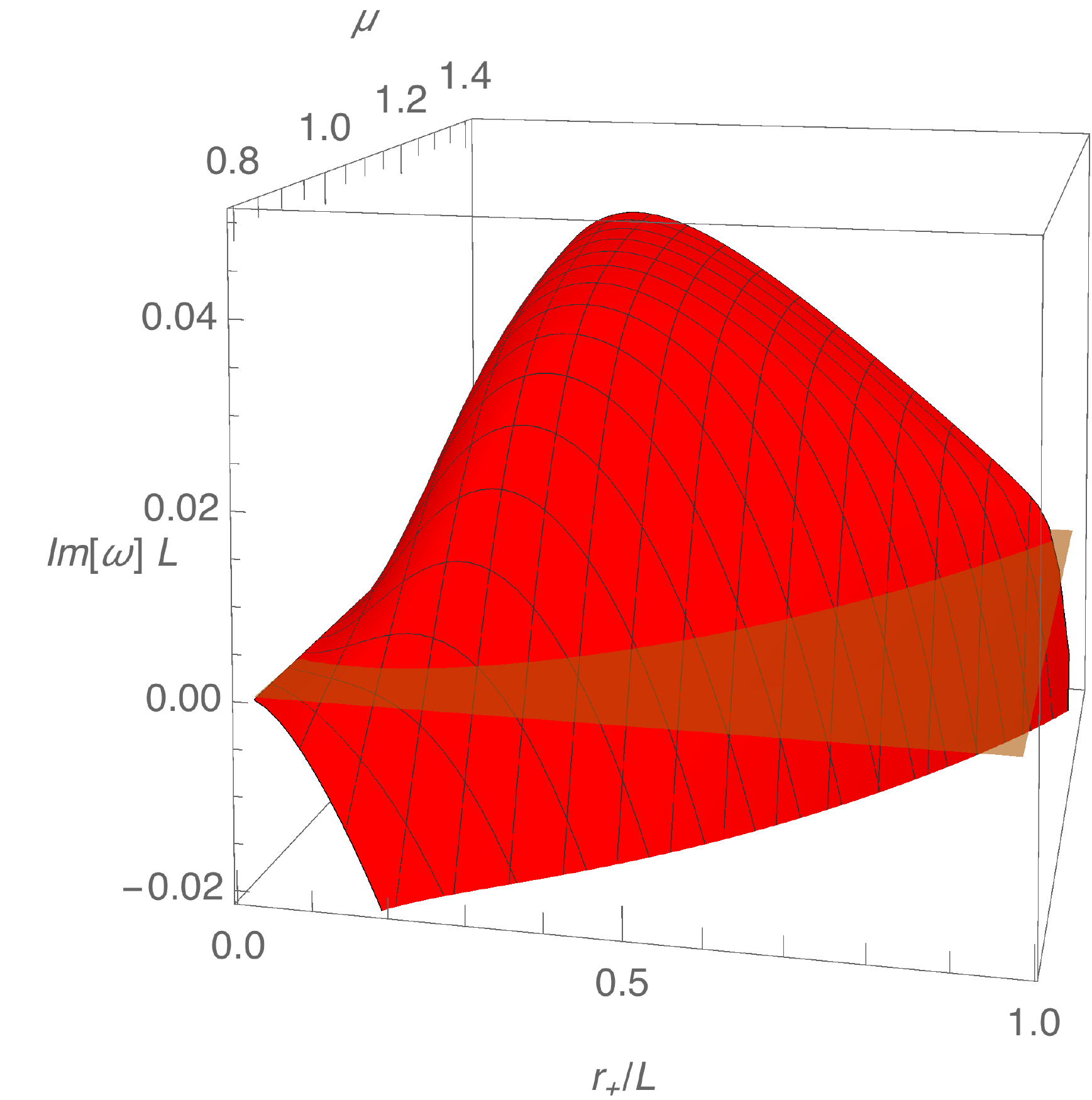}
\includegraphics[width=.48\textwidth]{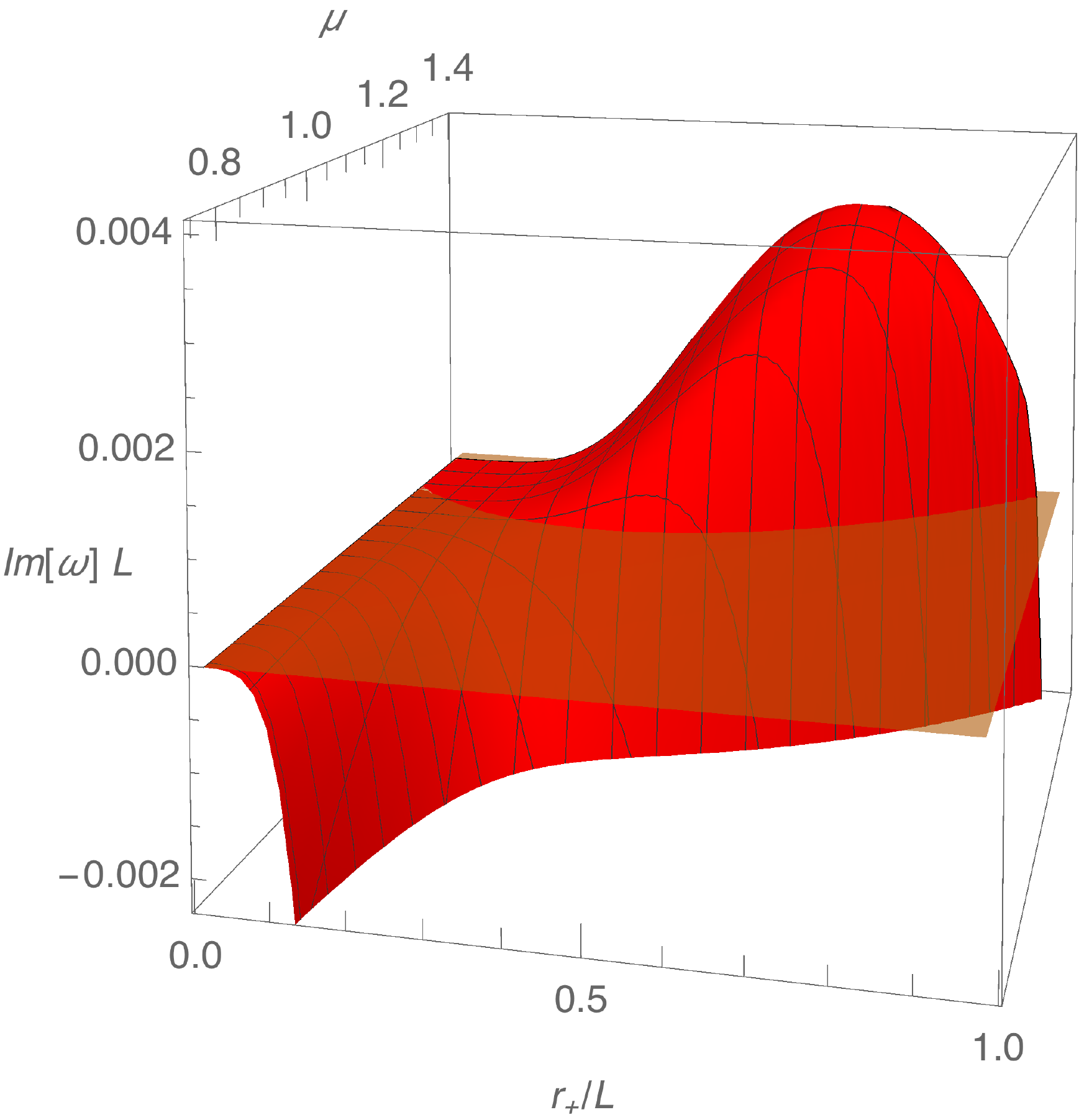}
}
\caption{The imaginary part of the frequency as a function of the chemical potential $\mu$ and the horizon radius $R_+$ for $e=3.5$. The ({\bf left panel}) is for the $\ell=0$ mode and the ({\bf right panel}) is for $\ell=1$.}
\label{fig:dominant}
\end{figure}

On the other hand,  properties 2) and 3) are illustrated, for $e=3.5$, in Fig. \ref{fig:dominant} which plots ${\rm Im}(\omega L)$ as a function of $R_+$ and $\mu$ for the $\ell=0$ mode  (left panel) and for the $\ell=1$ harmonic (right panel). The left panel is just a zoom in of the unstable region of Fig. \ref{fig:Re_Im_e}. The $\ell=0$ mode is indeed unstable in a wider range of phase space and its maximum ${\rm Im}(\omega L)$ is considerably larger. This is in agreement with the time domain analysis of \cite{Degollado:2013bha} (note that the frequency-domain study of \cite{Herdeiro:2013pia} just considered the $\ell=1$ harmonic). This trend continues for higher $\ell$. For $e=3.5$ we find that the maximum of the instability for the first three harmonics are 
\begin{eqnarray}
&& \ell=0: \:\: Max\{{\rm Im}(\omega L)\}\sim 4.0\times 10^{-2}, \nonumber \\
&& \ell=1: \:\: Max\{{\rm Im}(\omega L)\}\sim 3.3\times 10^{-3}, \nonumber \\
&& \ell=2: \:\: Max\{{\rm Im}(\omega L)\}\sim 6.3\times 10^{-5}. 
\end{eqnarray}
These properties are similar for other values of $e$. As a final example, for $e=1.9$ (the case whose $\ell=0$ data is plotted in Fig. \ref{fig:Re_Im_e1.9}) the maximum of the instability for the two first harmonics is
\begin{eqnarray}
&& \ell=0: \:\: Max\{{\rm Im}(\omega L)\}\sim 1.4\times 10^{-2}, \nonumber \\
&& \ell=1: \:\: Max\{{\rm Im}(\omega L)\}\sim 1.1\times 10^{-4}. 
\end{eqnarray}
In this case the instability shuts down for $\ell=2$ except on a very small corner around $\mu\sim \mu_{ext}$ and $R_+\sim 1$. At least for the values we considered, we find that the instability strength grows with $e$. 

\section{Non-interacting thermodynamic model for hairy black holes}\label{sec:noninteracting}

In previous sections we have seen that a RN BH with a charged scalar field confined inside a box is linearly unstable to the superradiant and near-horizon scalar condensation instabilities. This suggests that a RN BH inside a box should evolve to a different configuration that should have scalar hair (floating above the horizon and confined inside the box) that is stable to the original mechanisms that drive RN unstable. In the limit where the amplitude of the scalar hair vanishes, such hairy black hole should merge with the RN family at the onset of the instability.

 On the other hand, a scalar field inside a box in Minkowski space has its linear spectrum of frequencies quantized as we saw in Section \ref{subsec:Normal modes}.  Borrowing ideas from similar AdS and massive systems with confined potentials, it is natural to expect that such a normal mode of the scalar field can be back-reacted to higher order in perturbation theory (with the expansion parameter being the amplitude of the scalar field). That is to say, beyond linear order, the scalar field sources the gravitoelectric field but it should be possible to do so while keeping the solution asymptotically flat and regular everywhere (in particular, at the origin) and with the scalar hair confined inside the box. This should then yield a new horizonless solution of Einstein-Maxwell-scalar theory (with the confinement condition imposed) that is a boson star oscillating with frequency $ \tilde{\omega}$ (if we work in a gauge where the scalar field is complex with time dependence $e^{i \tilde{\omega} t}$) or, equivalently after a gauge transformation, a static soliton (with no time dependence and chemical potential $\mu= \tilde{\omega}/q$). 
 
 Again borrowing ideas from similar confined systems, it is also natural to expect that we can place a standard RN BH on top of this soliton to generate a hairy black hole. This hairy black hole family whose zero-horizon radius is the soliton of the theory should then be the same hairy black hole that connects (in a phase diagram of solutions) to the RN BH family at the onset of the instability discussed above.
 
Such hairy solitons and black holes would have a charged scalar condensate floating above the origin/horizon, with the electromagnetic repulsion balancing the gravitational collapse of the scalar condensate. For these ideas to be correct, in the case of the hairy black hole, we further expect that hairy black holes have higher entropy than the RN BH with the same mass and electric charge (when they coexist).

To prove these ideas correct, we have to solve the Einstein-Maxwell-scalar equations of motion in a confined box and find the proposed asymptotically flat solutions. We will do this in a companion publication \cite{DiasMasachs}. However, here we want to derive heuristically the {\it leading order} thermodynamic properties of such hairy solutions using a simple non-ineracting thermodynamic model that does {\it not} make use of the equations of motion. This model proved to capture the correct leading order physics of charged and rotating AdS systems  \cite{Basu:2010uz,Dias:2011at,Dias:2011tj,Cardoso:2013pza,Dias:2015rxy,Dias:2016pma}.

At leading order, a charged soliton (boson star) is just a normal mode of the ambient background (whose frequency $\tilde{\omega}$ is then corrected as we climb the perturbation expansion ladder). This is a 1-parameter family of solutions with mass ${\cal M}_{sol}=\mu\,{\cal Q}_{sol}+{\cal O}({\cal Q}^2)$ with $\mu=\frac{\tilde{\omega}}{q}$ (we are interested in the soliton with lowest energy which has  $\mu=\pi/e$). This soliton further obeys the first law, $d{\cal M}_{sol}=\mu \,d{\cal Q}_{sol}$. 

The non-interacting thermodynamic model of  \cite{Basu:2010uz,Dias:2011at,Dias:2011tj,Cardoso:2013pza,Dias:2015rxy,Dias:2016pma} assumes that we can place a small RN BH on top of this soliton to get a 2-parameter family of hairy BHs (the parameters being the mass and the charge). It further assumes that, at leading order (and certainly only at this order), the system can be considered as a {\it non}-interacting mixture of a soliton and a RN BH at its core in the sense that the mass and the charge of the hairy BH are given by the direct sum of the mass and charge of the RN BH and the soliton,
\begin{align} \label{masscharge}
{\cal M}={\cal M}_{sol}+{\cal M}_{RN} \quad \text{and} \quad {\cal Q}={\cal Q}_{sol}+{\cal Q}_{RN}. 
\end{align}
The soliton carries no entropy, so the entropy $S$ of the hairy BH simply reads
\begin{align} \label{entropy}
S=S_{RN}({\cal M}_{RN},{\cal Q}_{RN})+S_{sol}({\cal M}_{sol},{\cal Q}_{sol})=S_{RN}({\cal M}-{\cal M}_{sol},{\cal Q}-{\cal Q}_{sol}).
\end{align}

The hairy BH can partition its charge ${\cal Q}$ and mass ${\cal M}$ between the RN BH and soliton components. On physical grounds one expects this distribution to be such that, for fixed mass ${\cal M}$ and charge ${\cal Q}$, the entropy $S$ is maximised, $dS=dS_{RN}=0$, while respecting the first law of thermodynamics $\mathrm d {\cal M}=T_H \mathrm d S+\mu \mathrm d {\cal Q}$. The maximisation of the entropy turns out to imply \cite{Dias:2011tj,Dias:2016pma}
\begin{equation}\label{chemicaleq}
 \mu_{RN}=\mu_{sol}\equiv \mu \quad \hbox{and} \quad T_{RN}\equiv T_H,
\end{equation}
\ie, not surprisingly, the two mixed constituents must be in thermodynamic equilibrium to yield a hairy BH. 

The mass and charge of the soliton are related by ${\cal M}_{sol}=\mu {\cal Q}_{sol}$, while the mass and charge of the RN BH are given by \eqref{RNthermo}.
Using these relations together with the non-interacting relation \eqref{masscharge} and equilibrium conditions \eqref{chemicaleq}, we can express the leading order thermodynamic quantities of the hairy black hole (and of its two constituents) in terms of its mass ${\cal M}$, charge ${\cal Q}$ and chemical potential $\mu$:
\begin{align}\label{noninteracting}
\begin{split}
&R_+ =\frac{4 }{2-\mu ^2}({\cal M}-\mu  {\cal Q})+\mathcal O\left({\cal M}^2,{\cal Q}^2,{\cal M} {\cal Q}\right),\\
&T L=\frac{\left(2-\mu ^2\right)^2}{32 \pi  ({\cal M}-\mu  {\cal Q})}+\mathcal O\left({\cal M}^2,{\cal Q}^2,{\cal M} {\cal Q}\right),\\
&S/L^2=\frac{16 \pi }{\left(2-\mu ^2\right)^2} ({\cal M}-\mu  {\cal Q})^2+\mathcal O\left({\cal M}^3,{\cal Q}^3,{\cal M}^2 {\cal Q},{\cal M} {\cal Q}^2\right);\\
&\hspace{1cm} {\cal M}_{RN}=\frac{\left(2+\mu ^2\right)}{2-\mu ^2}({\cal M}-\mu  {\cal Q})+\mathcal O\left({\cal M}^2,{\cal Q}^2,{\cal M} {\cal Q}\right),\\
&\hspace{1cm} {\cal M}_{sol}=\frac{\mu  \left(2+\mu ^2\right) {\cal Q}-2 \mu ^2 {\cal M}}{2-\mu ^2}+\mathcal O\left({\cal M}^2,{\cal Q}^2,{\cal M} {\cal Q}\right),\\
&\hspace{1cm} {\cal Q}_{RN}=\frac{2 \mu }{2-\mu ^2} ({\cal M}-\mu  {\cal Q})+\mathcal O\left({\cal M}^2,{\cal Q}^2,{\cal M} {\cal Q}\right),\\
&\hspace{1cm} {\cal Q}_{sol}=\frac{\left(2+\mu ^2\right) {\cal Q}-2 \mu  {\cal M}}{2-\mu ^2}+\mathcal O\left({\cal M}^2,{\cal Q}^2,{\cal M} {\cal Q}\right).\\
\end{split}
\end{align}

The domain of existence of the hairy black hole can be  inferred from this analysis. In one extremum, the soliton constituent is absent and all the mass and charge of the hairy BH is carried by the RN component. This describes the hairy BH that merges with  the RN BH at the zero-mode of its linear instability. On the opposite extremum configuration, the  RN BH constituent is absent and the soliton component carries all the mass and charge of the solution. This is the zero-radius or zero-entropy limit of the hairy black hole. It follows that the hairy black hole mass must be within these two boundaries:
\begin{align}\label{bdriesThermoModel}
\mu  {\cal Q}+\mathcal O ({\cal Q}^2) \leq  {\cal M} \leq \frac{\mu ^2+2 }{2 \mu }{\cal Q}+\mathcal O ({\cal Q}^2).
\end{align}
Equating the two extrema configurations, yields an interval of existence for the chemical potential:
\begin{align}\label{condThermoModel}
\mu\leq\frac{\mu ^2+2}{2 \mu }\quad \Rightarrow  \quad \mu\leq\sqrt{2} \quad \implies e\geq \frac{\pi}{\sqrt{2}}\,, 
\end{align}
where in the last relation we used that for the ground state solution the leading order potential is $\mu=\tilde{\omega}/e=\pi/e +\mathcal O(R_+)$: see \eqref{eq:chemicalPotential} and footnote \ref{foot:Gauge}. The lower bound for $e$ in \eqref{condThermoModel} is precisely the leading term in \eqref{eq:onsetSuperradiance}  when we set the chemical potential to its extreme value, $\mu=\sqrt 2$.

The above leading order thermodynamic analysis must be considered with a few grains of salt. Indeed, first note that a theory can have hairy black holes that do not have a zero-radius limit, \ie$\text{ }$a solitonic limit (see \eg \cite{Dias:2011tj}). Second, there is no reason why the non-interacting mixture assumption \eqref{masscharge} should hold, even at leading order. Nevertheless, in a companion publication \cite{DiasMasachs}, we will solve the equations of motion to find the hairy BHs and we will confirm that the  thermodynamic model indeed captures the correct leading order thermodynamics \eqref{noninteracting}-\eqref{condThermoModel} of the system.

A final important observation is in order. To keep the scalar field of the asymptotically flat hairy black hole confined inside the box (\ie $\phi(R\geq 1)=0$), the latter must have a certain Lanczos-Darmois-Israel surface stress tensor 
\cite{Israel:1966rt,Israel404,Kuchar:1968,Barrabes:1991ng}. That is to say, the induced gravitoelectric fields are continuous at the thin layer surface that confines the scalar hair but the extrinsic curvature is discontinuous. The above thermodynamic model is completely blind to this information. It is too simple for that; yet it yields the correct leading thermodynamics.
In the aforementioned companion publication \cite{DiasMasachs}, we will emphasize the importance of using the Brown-York quasilocal formalism \cite{Brown:1992br} to appropriately discuss hairy black holes contained inside a box. For the difference between the  Brown-York tensor outside and inside the box surface layer yields precisely the Lanczos-Darmois-Israel  surface stress tensor \cite{Israel:1966rt,Israel404,Kuchar:1968,Barrabes:1991ng} that describes the energy-momentum content of the box \cite{Brown:1992br}. 
The leading order thermodynamic analysis that we do in the present manuscript is valid for $R_+\ll 1$, that is to say, for  small mass and charge BHs. In this regime, the ADM mass $M$ and charge $Q$ are the same as the Brown-York quasilocal mass ${\cal M}$ and charge ${\cal Q}$.\footnote{This also explains why the thermodynamic model is blind to the presence of the box: the regime $r_+\ll L$ can be understood as the limit where the box radius $L$ is large. In the limit $L\to \infty$, the Brown-York charges reduce to the ADM ones and the presence of the box goes unnoticed.} Yet, this is no longer the case at higher orders. For this reason, and to keep notational consistency with the discussions of the companion publication, in the analysis of this section we have used the calligraphic letters ${\cal M}$ and ${\cal Q}$ when referring to the mass and charge of the solution. 

\section{Conclusions}
 
 In this manuscript we have considered the Reissner-Nordstr\"om$-$mirror system, aka charged black hole bomb system. We pointed out that this system is unstable not only to superradiance but also to the near-horizon scalar condensation instability. These two observations and some numerical analysis allow to identify with great accuracy the black hole and scalar field parameters that are required to have instability. This should be of valuable use to choose initial data in Cauchy studies of the evolution of the instability. 
From our findings and from \cite{Dias:2010ma}, it follows that the near-horizon scalar condensation instability should co-exist with the superradiant instability also in the original Kerr black hole bomb system of Press and Teukolsky \cite{Press:1972zz}. This, and other implications of our observation are currently under investigation. 

 We also studied the properties, namely the quantized spectrum and growth rates, of these instabilities in detail. Finally, we pointed out that, in a phase diagram of solutions of the theory,  the onset of the instability should signal a branching-off to a new family of solutions that describe charged hairy black holes with scalar hair floating above the horizon. Electric repulsion should balance the system against gravitational collapse. Without using the equations of motion, we can predict the leading order thermodynamics of these charged hairy black holes. For that, we used a non-interacting thermodynamic model that assumes that, at leading order in perturbation theory, the hairy black hole can be seen as a small Reissner-Nordstr\"om black hole placed at the core of the caged soliton of the theory, as long as these two constituents are in thermodynamic equilibrium (\ie if they have the same chemical potential). 
 
 Confirmation that these hairy solutions do indeed exist beyond linear order in perturbation theory will be given in the companion manuscript \cite{DiasMasachs}, where we explicitly find these solutions solving the equations of Einstein-Maxwell theory with a complex scalar field that is confined inside a box. We will find that our hairy black holes always have higher entropy than caged Reissner-Nordstr\"om black holes with the same energy and electric charge. Therefore, they are the natural candidates for the endpoint of charged superradiance in the Reissner-Nordstr\"om$-$mirror system.
In previous literature boxed hairy black holes were already considered. However these studies differ from ours in two main aspects. The hairy black holes of \cite{Dolan:2015dha,Ponglertsakul:2016wae,Ponglertsakul:2016anb} are  regular in the inner boundary but the gravitoelectric fields diverge logarithmically in the asymptotic region: they are not asymptotically flat. On the other hand, Ref.  \cite{Basu:2016srp} focused its attention on describing the solution in the interior of the box. In  \cite{DiasMasachs} we will describe the hairy black hole in the interior and exterior regions. In particular, this will require studying the Israel junction conditions at the location of the box and discuss the surface stress tensor that this surface layer must have to obey the energy conditions as required in a physical system.

\vskip .5cm
\centerline{\bf Acknowledgements}
\vskip .2cm
We wish to thank Ian Hawke, Jorge Santos, Kostas Skenderis and Marika Taylor for discussions. 
The authors acknowledge financial support from the STFC Ernest Rutherford grants ST/K005391/1 and ST/M004147/1. OD further acknowledges support from the STFC ``Particle Physics Grants Panel (PPGP) 2016", grant ref. ST/P000711/1. 


\begin{appendix}

\section{No-hair theorem (in the absence of a confining box)}\label{App:HoHair}

In the original spirit of black hole bomb systems, the idea is to have a box that confines a scalar field in a black hole background \cite{Press:1972zz}. This is certainly possible at linear level and we might conjecture that it should then be possible to extend this solution to higher order in perturbation theory and at full nonlinear level. Such black holes should be asymptotically flat (like the original unstable black hole) and regular everywhere. This should be possible when the scalar hair is totally confined inside the box, \ie when it vanishes outside it. These are the kind of solutions that will be found in \cite{DiasMasachs}, and already partially discussed in \cite{Basu:2016srp}, and whose leading order thermodynamics is anticipated in Section \ref{sec:noninteracting} using a simple thermodynamic model.

Yet, in \cite{Dolan:2015dha,Ponglertsakul:2016wae,Ponglertsakul:2016anb} a distinct solution with scalar hair was constructed numerically. Within Einstein-Maxwell-scalar theory, the hairy solitonic and black hole solutions of  \cite{Dolan:2015dha,Ponglertsakul:2016wae,Ponglertsakul:2016anb}  have the following properties: 1) they are regular at the origin or horizon, 2) the scalar field is required to vanish at a given timelike hypersurface, $R=1$, {\it but not outside it}, and  3) the solution is {\it not} asymptotically flat.   

In this appendix, we explain why the hairy solutions of  \cite{Dolan:2015dha,Ponglertsakul:2016wae,Ponglertsakul:2016anb} are necessarily not asymptotically flat. 
For that, we construct these solutions in a perturbative expansion. While doing so,  we will effectively formulate a no-hair theorem for asymptotically flat regular hairy solutions whose scalar field is not totally confined inside a box. A box in these conditions is not particularly relevant and, if we drop it, we effectively recover the original no-hair theorems of   \cite{Ruffini:1971bza,Chrusciel:1994sn,Bekenstein:1996pn,Heusler:1998ua}: {\textit{The only asymptotically flat, spherically symmetric and static (regular) black hole solutions of Einstein-Maxwell theory are members of the Reissner-Nordstr\"om family.}}

Following \cite{Dolan:2015dha,Ponglertsakul:2016wae,Ponglertsakul:2016anb}, we set $\phi(R)\big|_{R=1}=0$. We fix the expansion parameter $\varepsilon$ of our perturbation theory to be the first derivative of the scalar field at the box,
\begin{align}
&\phi(R)|_{R=1^-}=\varepsilon (1-R) +\mathcal O\left((1-R)^2\right), \nonumber \\
&\phi(R)|_{R=1^+}=-\varepsilon (R-1) +\mathcal O\left((R-1)^2\right).
\end{align}
Also following \cite{Dolan:2015dha,Ponglertsakul:2016wae,Ponglertsakul:2016anb}, we further require that the gravitational, Maxwell and scalar fields are regular at the origin or horizon (but we do not require the solutions to be asymptotically flat).

In these conditions we look for an hairy soliton that has the perturbative  expansion:
\begin{align}\label{eq:schemeSoliton}
\begin{split}
&f(R)=1+ \sum_{n\geq 1}\varepsilon^{2n}f_{2n}(R),\\
&A(R)=\frac{\pi}{e}+\sum_{n\geq 1}\varepsilon^{2n}A_{2n}(R),\\
&\phi(R)=\sum_{n\geq 0}\varepsilon^{2n+1}\phi_{2 n+1}(R).
\end{split}
\end{align}
The leading order Maxwell potential is $\pi/e$ because, after a $U(1)$ gauge transformation, this corresponds to pick the normal mode with lowest frequency as a linear seed for the scalar field. 

At  order $\mathcal O(\varepsilon)$ we have to solve the Klein-Gordon equation for the scalar field $\phi_1$ about the Minkowski spacetime. The most general solution is:
\begin{align}
\phi_1(R)=\frac{e^{-i \pi  R}}{2 R} \left(2 \beta_1-\frac{i \beta_2 e^{2 i \pi  R}}{\pi }\right).
\end{align}
Regularity at the origin implies $\beta_2=-2 i \pi  \beta_1$, while at the two conditions at the  box fix $\beta_1=-\frac{i \varepsilon}{2 \pi }$.

At order $\mathcal O(\varepsilon^2)$, the $\mathcal O(\varepsilon)$ scalar field sources a back-reaction on the gravitoelectric fields. Solving the equations of motion for $A_2$ and $f_2$ we obtain:
\begin{align}
&A_2(R)=C_2^{A_2}+\frac{e \sin (\pi  R) \cos (\pi  R)-\pi  (\pi  C_1^{A_2}+e R [\text{Ci}(2 \pi  R)-\log R])}{\pi ^2 R},\nonumber \\
&f_2(R)=C_2^{f_2}-\frac{C_1^{f_2}}{R}-2 \text{Ci}(2 \pi  R)+2 \log R+\frac{\sin (2 \pi  R)}{\pi  R}.
\end{align}
We can fix two integration constants by imposing regularity at the origin: $C_1^{f_2}=C_1^{A_2}=0$. A series expansion for large $R$ then yields:
\begin{align}\label{LogDiv}
&A_2(R)|_{R\to \infty}\simeq C_2^{A_2}+\frac{e \cos (2 \pi  R)}{4 \pi ^3 R^2}+\frac{e  \log R}{\pi }, \nonumber \\
&f_2(R)|_{R\to \infty}\simeq C_2^{f_2}+\frac{\cos (2 \pi  R)}{2 \pi ^2 R^2}+2  \log R.
\end{align}
We see that the two fields diverge as the $ \log R$, and we cannot fix any integration constant  to eliminate these divergences. These logarithmic divergences are sourced by the $\mathcal O(\varepsilon)$ scalar field (that is to say, the two homogeneous solutions of the  $\mathcal O(\varepsilon^2)$ equations of motin behave as a constant and $1/R$ as $R\to \infty$). This is a consequence of not requiring that the scalar field vanishes outside the box. 

Hairy black hole solutions can be constructed using a similar perturbative scheme, this time using a double parameter expansion ($\varepsilon$ and the horizon radius $R_+$). Thus, with the conditions proposed by \cite{Dolan:2015dha,Ponglertsakul:2016wae,Ponglertsakul:2016anb}, the black hole fields will have the same  logarithmic divergence observed in \eqref{LogDiv}.

The logarithmic divergence of the fields already at order $\mathcal O(\varepsilon^2)$ explains why the solitons and hairy black holes of \cite{Dolan:2015dha,Ponglertsakul:2016wae,Ponglertsakul:2016anb} are not asymptotically flat. 
Note that `no-hair' theorems typically have two golden assumptions: 1) the solutions should be regular everywhere and, in particular, at the inner boundary, and 2) the solutions should be asymptotically flat  \cite{Ruffini:1971bza,Bekenstein:1996pn}. The solutions of \cite{Dolan:2015dha,Ponglertsakul:2016wae,Ponglertsakul:2016anb} (further discussed in \cite{Sanchis-Gual:2016tcm,Sanchis-Gual:2016ros}) violate the latter assumption so it comes without surprise that they also evade the `no-hair' theorems. 

In a companion publication \cite{DiasMasachs}, we will however find hairy solitons and black holes that are regular everywhere and asymptotically flat. For that, we will follow strictly the spirit of the charged black hole bomb and require that the scalar field vanishes at, but also outside, the box. The leading order thermodynamics of such hairy solutions is already given in Section \ref{sec:noninteracting}.

\section{Numerical convergence tests}\label{App:Numerics}

For the numerics in Section \ref{sec:QNM} we have used both a generalised eigenvalue problem scheme (see Section III.C in \cite{Dias:2015nua}) and a Newton-Raphson scheme (see Section III.C and VI.A in \cite{Dias:2015nua}). 

\begin{figure}[t]
\centerline{
\includegraphics[width=.48\textwidth]{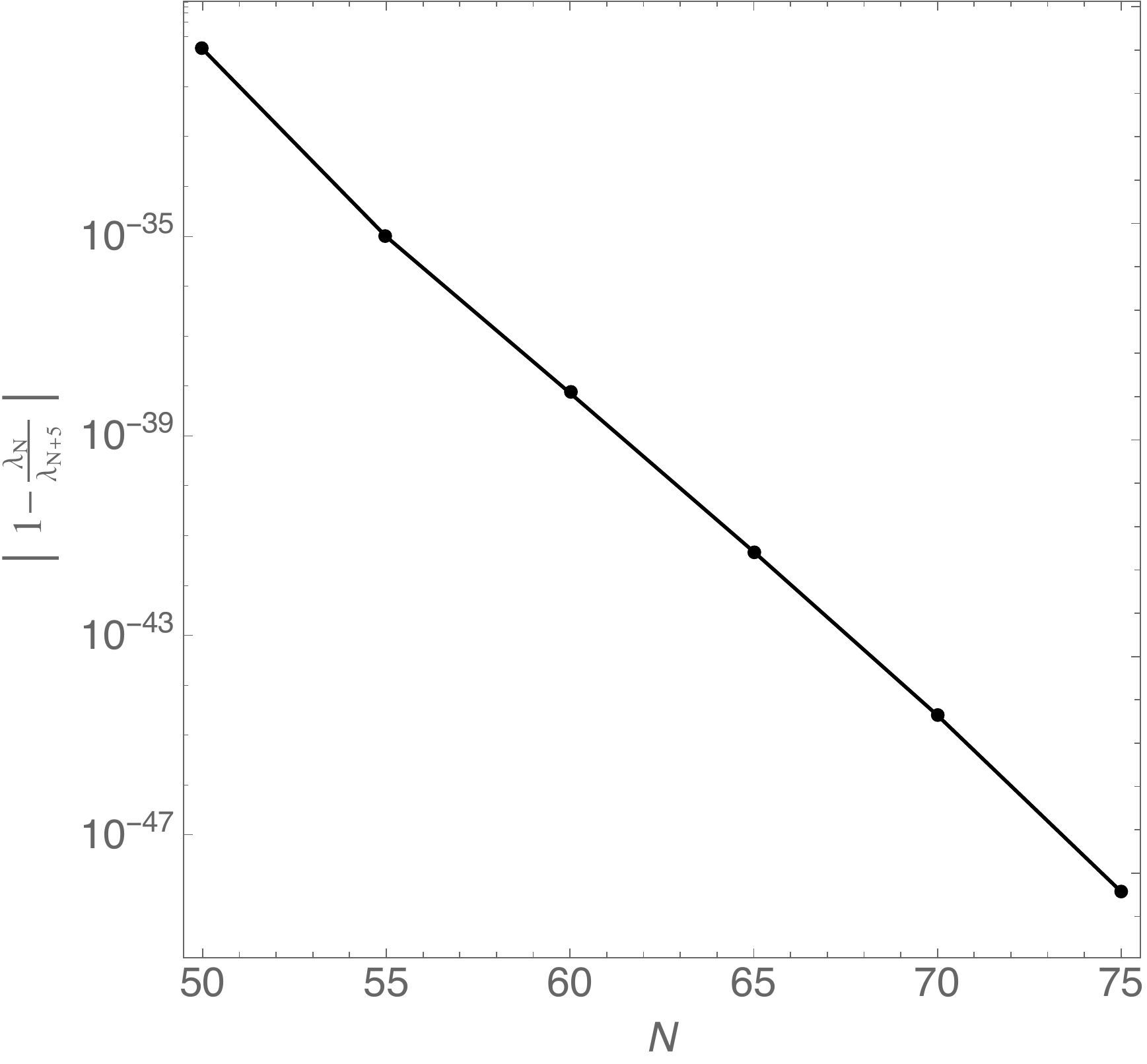}
\hspace{0.3cm}
\includegraphics[width=.48\textwidth]{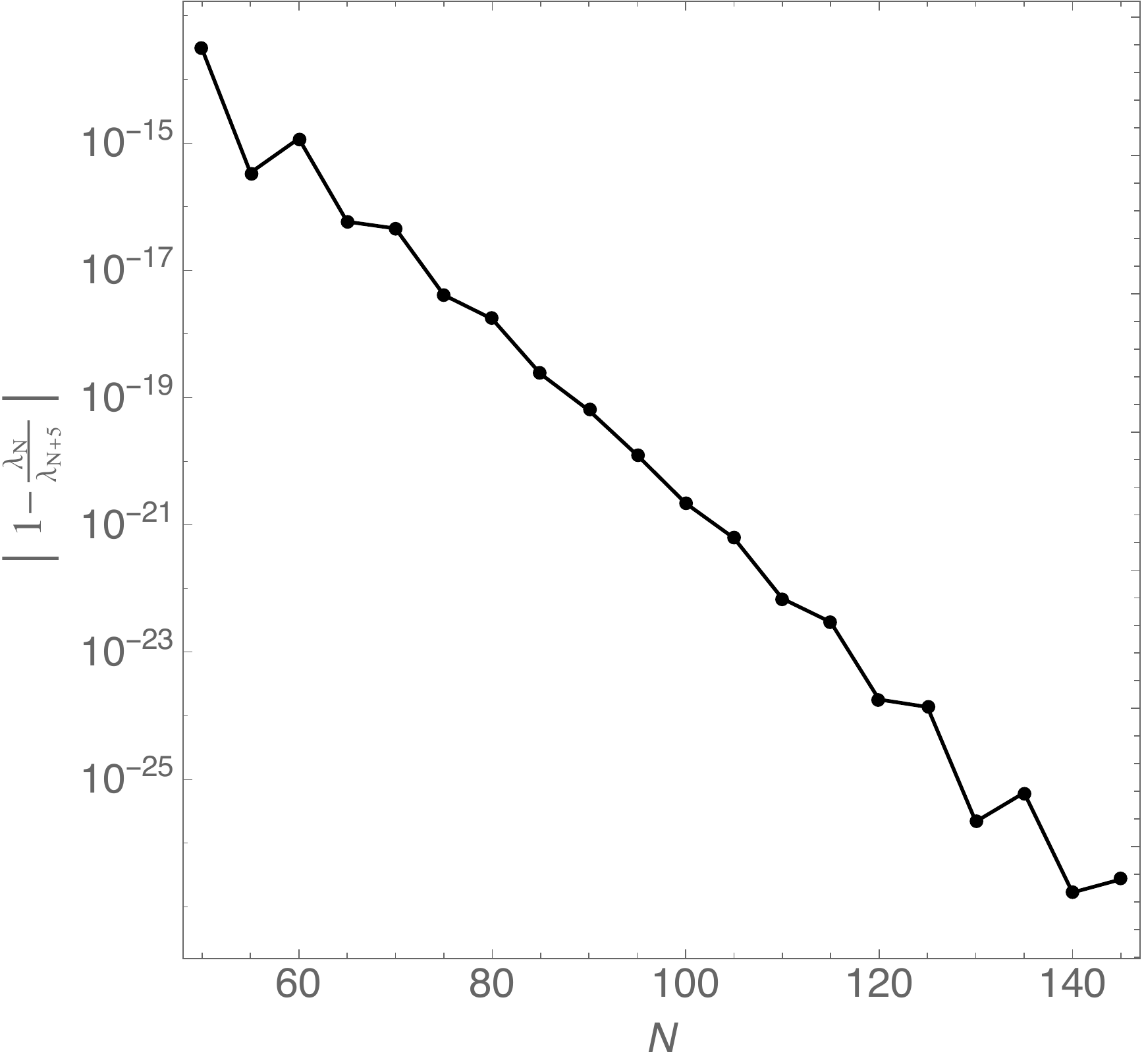}
}
\caption{Illustration of convergence tests. For this, we focus on the curve with chemical potential $\mu=\sqrt 2 (1-10^{-1})$ in Fig. \ref{fig:eonset_Rp_chp}. Then we study convergence for the solution with $R_+=0.4$ (left panel), and with $R_+=0.98$ (right panel). The latter, being closer to the box radius requires higher resolution to achieve similar accuracy. Higher resolution is also typically required as we approach extremality.}
\label{fig:ConvergenceTest_onset}
\end{figure}

To determine the number of points in the Chebysehv grid that are required to get accurate results, we have done convergence tests to find the number of grid points $N$ that we should use and to confirm that our numerical code is delivering the exponential convergence as required when pseudospectral discretization is used.  
In order to measure the accuracy of the numerical results in Section \ref{sec:QNM} we define the error as a function of the number of points $N$ as:
\begin{align}
\text{err}_N=\left|1-\frac{\lambda_{N}}{\lambda_{N+\Delta N}}\right|,
\end{align}
where $\lambda_N$ is the eigenvalue obtained with a grid of $N$ points and $\Delta N$ is the increase of the number of points; in the examples of Fig. \ref{fig:ConvergenceTest_onset}, $\Delta N=5$.

One of the convergence tests for the onset of instability Fig. \ref{fig:eonset_Rp_chp} is shown in Fig. \ref{fig:ConvergenceTest_onset}. We have done similar tests for various horizon radius and determined the number of grid points that satisfies the error bounds we wish: Our $e_{onset}$'s and $\omega$'s are accurate up to at least the $8^{\rm th}$ decimal digit. For example, we required an error smaller than $10^{-10}$ for the curves with chemical potential $\mu=\sqrt 2 (1-10^{-5})$. $N= 81$ is enough to fulfill this requirement. However, in order to satisfy the same error for chemical potentials nearer to extremality we have increased the number of points to $N=251$ in Fig. \ref{fig:eonset_Rp_chp} for $\mu=\sqrt 2 (1-10^{-n})$ with $n>5$.
As an example of our convergence tests, in the left panel of Fig. \ref{fig:ConvergenceTest_onset} we display the error as a function of the number of grid points $N$ for the chemical potential $\mu=\sqrt 2 (1-10^{-1})$ and horizon radius $R_+=0.4$. The error follows the expected exponential decay for pseudospectral methods in a Chebyshev grid. On the right plot we display the same quantities but this time for the chemical potential $\mu=\sqrt 2 (1-10^{-1})$ and horizon radius $R_+=0.98$. This curve has higher error, as expected, since we are approaching the limiting case $R_+=1$. 

We have performed convergence tests also for the frequency computations of Section \ref{sec:QNM}. However, in general the most challenging computations were those associated to Fig. \ref{fig:eonset_Rp_chp} (hence our choice of presenting the convergence tests for this computation). This is expected since in this case  we had to consider the most extreme situations: we had to approach extremality by tuning the chemical potential and also the limiting cases $R_+=0$ and $R_+=1$.

\end{appendix}


\bibliography{refs_BoxQN}{}
\bibliographystyle{JHEP}

\end{document}